\documentclass[fleqn,usenatbib]{mnras}

\usepackage[T1]{fontenc}

\DeclareRobustCommand{\VAN}[3]{#2}
\let\VANthebibliography\thebibliography
\def\thebibliography{\DeclareRobustCommand{\VAN}[3]{##3}\VANthebibliography}

\usepackage{graphicx}	
\usepackage{amsmath}	
\usepackage{amssymb}	
\usepackage{newtxtext,newtxmath}
\usepackage[dvipsnames]{xcolor}

\newcommand{\diff}{\ensuremath{\mathrm{d}}}
\newcommand{\deriv}[2]{\ensuremath{\frac{\diff #1}{\diff #2}}}
\newcommand{\pderiv}[2]{\ensuremath{\frac{\partial #1}{\partial #2}}}

\title[Catastrophically evaporating planets]{Dust formation in the outflows of catastrophically evaporating planets}

\author[R. A. Booth et al.]{
Richard A. Booth,$^{1}$\thanks{E-mail: r.booth@imperial.ac.uk}
James E. Owen$^{1}$
and Matth\"aus Schulik$^{1}$
\\
$^{1}$Astrophysics Group, Department of Physics, Imperial College London, Prince Consort Rd, London, SW7 2AZ, UK\\
}

\date{Accepted 2022 October 26. Received 2022 October 26; in original form 2022 April 26}

\pubyear{2022}

\begin{document}
\label{firstpage}
\pagerange{\pageref{firstpage}--\pageref{lastpage}}
\maketitle

\begin{abstract}
Ultra-short period planets offer a window into the poorly understood interior composition of exoplanets through material evaporated from their rocky interiors. Among these objects are a class of disintegrating planets, observed when their dusty tails transit in front of their host stars. These dusty tails are thought to originate from dust condensation in thermally-driven winds emanating from the sublimating surfaces of these planets. Existing models of these winds have been unable to explain their highly variable nature {and have not explicitly modelled} how dust forms in the wind. Here we present new radiation-hydrodynamic simulations of the winds from these planets, including a {minimal} model for the formation and destruction of dust, {assuming that nucleation can readily take place}. We find that dust forms readily in the winds, a consequence of large dust grains obtaining lower temperatures than the planet's surface. {As hypothesised previously}, we find that the coupling of the planet's surface temperature to the outflow properties via the dust's opacity can drive time-variable flows when dust condensation is sufficiently fast. In agreement with previous work, our models suggest that these dusty tails are a signature of catastrophically evaporating planets that are close to the end of their lives. Finally, we discuss the implications of our results for the dust's composition. {More detailed hydrodynamic models that self-consistently compute the nucleation and composition of the dust and gas are warranted in order to use these models to study the planet's interior composition.}
\end{abstract}

\begin{keywords}
planets and satellites: atmospheres -- planets and satellites: surfaces
\end{keywords}



\section{Introduction}

While the occurrence, period, and radius distribution of short-period exoplanets has been extensively categorised through surveys with transit and radial velocity methods, the nature of exoplanet rocky interiors remains poorly understood. In particular, inferences about the composition, structure, and evolution of exoplanet interiors derived from mass-radius relations  are plagued by weak constraints and degeneracies \citep[e.g.][]{Valencia2007,Rogers2010,Dorn2017}. However, recent constrains derived from photoevaporation modelling of the `radius gap' \citep{Fulton2017} favour a broadly Earth-like composition \citep{Owen2017,Wu2019,Rogers2021}. Beyond density measurements, our best guesses for the composition of exoplanet rocky interiors are derived from extrapolations from the solar system planets. Recently, polluted white dwarf stars have begun to offer insights into plausible interior compositions \citep[e.g.][]{Gansicke2012,Hollands2021,Harrison2021}; however, the nature of the bodies probed, such as their size or formation location, is uncertain.

Ultra-short period planets offer a direct window into exoplanet interiors. These planets receive intense stellar radiation that can heat their surfaces to $\gtrsim 2000\,{\rm K}$. Under such conditions any primordial atmosphere would be lost through photoevaporation \citep[e.g.][]{Valencia2010}, leaving their surfaces exposed to the stellar radiation. As a result the surface will be molten on their day sides and gasses may escape to form rock-vapor atmospheres that consist predominantly of Na, O, O$_2$, and SiO, along with Fe and Mg \citep{Schaefer2009, Miguel2011, Kite2016}.

For earth-mass planets and above, these high mean-molecular-weight atmospheres might be observed through molecules such as SiO or SiO$_2$ \citep{Ito2015,Zilinskas2022}.  Efforts to observe these atmospheres have so far been largely inconclusive. For example, the nature of 55 Cnc e's atmosphere remains heavily debated, with the large hotspot shift seen in its thermal phase curve  \citep{Demory2016} not currently explained by atmosphere models \citep{Hammond2017,Kipping2020,Morris2021}. Recently, however, LHS 3844b has been shown not to harbour a massive atmosphere \citep{Kreidberg2019}, and it has been suggested that the phase curve of K2-141b is consistent with a rock vapor atmosphere \citep{Zieba2022}. The nature of these atmospheres will likely become clear soon, when the planets are observed by the \emph{James Webb Space Telescope} \citep{Zilinskas2022}.

Lower-mass planets, below roughly the mass of mercury ($\sim0.05\,M_\oplus$), instead undergo extreme mass loss, potentially being destroyed within a few Gyr \citep[e.g.][]{PB2013}. Indeed, the existence of these disintegrating planets was first inferred from the \emph{Kepler} light curve of KIC 12557548b (Kepler-1520b, \citealt{Rappaport2012}; see \citealt{vanLieshout2018} for a review). Two other systems, KOI-2700b \citep{Rappaport2014} and K2-22b \citep{Sanchis-Ojeda2015} have since been discovered. The light curves show dips in brightness with a distinct period but depths that vary from transit to transit. Together with a small pre-transit brightening that is well explained by forward scattering from dust grains, it is clear that these objects are best described by elongated tails of dusty material that has escaped from a planet. The size of the dust grains produced in these outflows has been estimated a number of ways. Constraints from the forward scattering peaks, and a possible important role of radiation pressure suggest grain sizes of order $0.1$ to 1~\micron{} \citep{Brogi2012,Budaj2013,Sanchis-Ojeda2015}, while the grey absorption spectrum of K2-22b suggests sizes $\gtrsim 0.5\,\micron{}$ \citep{Schlawin2021}.

The physics of the outflows and the formation of dust within them remains debated, however. \citet{Rappaport2012} and \citet{PB2013} recognized that the gaseous outflows from low mass objects would be largely similar to the sublimation-driven outflows from comets, but that the origin of dust within them is expected to be different -- in comets the dust is simply entrained, but for the disintegrating planets dust must condense in the outflows. \citet{PB2013} showed that outflows driven by vaporization of the planet's surface could produce the required mass-loss rates, if the planets are small enough. However, they did not model dust formation, but parameterized it {and argued that dust should form because the gas can become supersaturated.}

Subsequently, \citet{Ito2015} and \citet{Kang2021} suggested that outflows from the day side of these planets should be dust free. \citet{Ito2015} based this on atmosphere models that predicted atmosphere temperatures that were too high to form dust. {In models with a more complete set of opacities, \citet{Zilinskas2022} recently found a range of behaviours depending on the atmospheric composition, including lower temperatures for atmospheres that are depleted of sodium, as might be expected for planets that have lost a non-negligible amount of mass \citep{Schaefer2009}}. \citet{Kang2021} suggested that dust formation in winds from the night side might more naturally explain the tails of disintegrating planets since dust lost from the planet's night side will have a larger orbital period. Strong flows from the day side of the planet to night side are expected \citep{Castan2011,Nguyen2020}, due to the large temperature differential between the two hemispheres that arises because the planets are tidally locked \citep[e.g.][]{Winn2018}. {Such day-night side transport has been seen in simulations of evaporating hydrogen dominated atmospheres in both 2D \citep{ProgaStone} and 3D \citep{Tripathi2015}; however multi-dimensional simulations are required to explore day-to-night side flows in the rock vapour case since condensation could substantially modify the dynamics.}  

However, of the three known disintegrating planets, only two of them show trailing dusty tails, while the third, K2-22b, shows a symmetric transit. Further, K2-22b has a forward scattering peak that occurs after the transit instead of before \citep{Rappaport2014, Sanchis-Ojeda2015, vanLieshout2018}, suggesting that it has a leading tail. The trailing tails of Kepler-1520b and KOI-2700b have instead typically been interpreted in terms of the role of radiation pressure in the tails dynamics \citep[e.g.][]{vanLieshout2014,Ridden-Harper2018}, rather than mass loss from the night side. In the absence of radiation pressure, outflows launched from the planet's day side produce leading tails, while radiation pressure increases the dust's orbital period, leading to trailing tails.  It has however not yet been possible to explain a leading tail from dust produced in winds emanating from the planet's night side.

In this paper we revisit the physics of mass loss from the planet's day side. To this end we have developed a 1D radiation-hydrodynamic model of outflow from the day side. For the first time, we include a simple model for dust condensation, showing that under appropriate conditions condensation can occur. {However, our dust model remains simplified, in that we do not explicitly treat the formation of seed nuclei. Instead, we assume that sufficient nuclei can form, justifying this because the gas becomes supersaturated. This model is a stepping stone towards a more complete model that can model the formation of seed nuclei and their subsequent evolution in the future. } A second key difference to previous models is that we employ a detailed {radiative transfer} calculation of the gas and dust temperatures, which {demonstrates that} the dust temperature is able to drop below the planet's surface temperature. This lower dust temperature promotes dust formation, as discussed in \autoref{sec:fiducial}.

In \autoref{sec:model} we describe our model used to explore dust formation in these winds. Next (\autoref{sec:results}), we present the results of our simulations, explaining requirements for dust formation and discussing the conditions in which they arise. In \autoref{sec:unsteady} we show that some of our models produce time-variable behaviour and discuss the conditions that give arise to it. Finally, in \autoref{sec:discussion} and \autoref{sec:conclusions} we discuss our results and present our conclusions.

\section{Model}
\label{sec:model}

We model the mass-loss from the sub-stellar point using a 1D spherically-symmetric hydrodynamic model in which the dust is treated as a fluid. The model couples gas dynamics, dust formation and dynamics, and radiative transfer. The equations solved for each species, $s \in \{\rm g, d\}$ (where $\rm g$ and $\rm d$ refer to gas, i.e. evaporated rock vapor, and dust, respectively), are
\begin{align}
    \pderiv{\rho_s}{t} + \nabla \cdot [\rho_s u_s] &= Q_s, \label{eqn:hydro_mass} \\
    \pderiv{\rho_s u_s}{t} + \nabla \cdot [\rho_s u_s^2 + P_s] &= - \rho_s \nabla \Phi + \deriv{\mathcal{M}_s}{t} + Q'_s, \\
    \pderiv{E_s}{t} + \nabla \cdot [u_s (E_s + P_s)] &= - u_s \rho_s \nabla \Phi \nonumber \\
    & \quad + u_s \deriv{\mathcal{M}_s}{t} + \deriv{\mathcal{E}_s}{t} + Q''_s + \Gamma_s, \label{eqn:hydro_en}
\end{align}
which are standard for photoevaporation models \citep{Schulik2022}. Here, $\rho_s$, $u_s$, and $P_s$ are the density, radial velocity and partial pressure of the species, respectively. Similarly, we denote the corresponding temperature of each species by $T_s$. The total energy, $E_s = \tfrac{1}{2}\rho_s u_s^2 + e_s$, where $e_{\rm s}$ is the internal energy. {We model $u_s = C_{{\rm V},s} T_s$, where the specific heat capacity. For the gas, the heat capacity is taken to be that of an ideal gas with $\gamma=1.28$ {with a mean molecular weight appropriate for the composition described in Appendix~\ref{app:Gas_Opac}}, $\mu=30\,{\rm amu}$, while for the dust we assume the high temperature Debye limit, i.e. $3k_{\rm B}$ per atom  where $k_{\rm B}$ is the Boltzmann constant.}  While $P_{\rm d}$ is negligible {due to the large mass of the dust grains}, it is formally included {as the pressure of an ideal gas of dust particles with kinetic temperature, $T_{\rm d}$}. The terms $Q_s, Q'_s$, and $Q''_s$ refer to the mass, momentum and energy source terms due to dust formation, while $\Gamma_s$ is the net radiative heating/cooling, and
\begin{equation}
\Phi = -\frac{GM_{\rm p}}{r} + 3\frac{GM_{*}}{a_{\rm p}^3} r^2    
\end{equation}
is the potential including the tidal term, and $M_{\rm p}$, $M_{\rm *}$ are the planet and stellar mass, $G$ is the gravitational constant, $a_{\rm p}$ is the planet's semi-major axis and $r$ is the distance from the planet.

The momentum coupling term (drag) is computed assuming the low Mach Number limit of the Epstein drag law, 
\begin{equation}
    \deriv{\mathcal{M}_{\rm d}}{t} = - \deriv{\mathcal{M}_{\rm g}}{t} = 
        \frac{\rho_{\rm d}\rho_{\rm g}}{\rho_{\rm DUST}} \frac{\bar{v_{\rm t}}}{a} (u_{\rm g} - u_{\rm d}) \label{eqn:coll_mom},
\end{equation}
where $\rho_{\rm DUST} = 3{\rm g\,cm}^{-3}$ is the internal density of the dust grains and $a$ is their radius. $\bar{v_{\rm t}} = \sqrt{8k_{\rm B} T_{\rm g}/\upi\mu}$, is the r.m.s. speed of the gas. The Epstein drag law is valid for particles that are small compared with the mean free path of the gas, which is always true for the systems considered \citep[e.g][]{PB2013}.

The energy coupling term,
\begin{align}
    \deriv{\mathcal{E}_{\rm d}}{t} &=\frac{\rho_{\rm d}\rho_{\rm g}}{\rho_{\rm DUST}} \frac{\bar{v_{\rm t}}}{a} \frac{1}{m_{\rm d} + m_{\rm g}} \left[ m_{\rm g} (u_{\rm g} - u_{\rm d})^2 + 3 k_{\rm B} (T_{\rm g} - T_{\rm d}) \right], \label{eqn:coll_en} \\
     &= \dot{\mathcal{E}}_{1,\rm d}+ \dot{\mathcal{E}}_{2, \rm d}, \nonumber 
\end{align}
includes two contributions: the heat generated by friction (the first term in the square brackets, $\dot{\mathcal{E}}_{1,\rm d}$)  and the exchange of thermal energy via collisions (the second term, $\dot{\mathcal{E}}_{2,\rm d}$). The equivalent expressions for the gas is obtained by swapping the subscripts, d and g.

\subsection{Dust formation}

We use a simplified model of dust formation and sublimation, designed to capture the appropriate equilibrium and approximate time-scale of evolution without modelling the chemistry of the gas and dust in detail. The growth rate of a dust grain due to condensation may be approximated by the collision rate between the grain and the gas,
\begin{equation}
    \left. \deriv{m}{t}\right|_{\rm cond} = \upi a^2 \alpha n_{\rm g} \bar{v_{\rm t}} \mu = 4 \upi a^2 \alpha \frac{P_{\rm g}}{\sqrt{2\upi k_{\rm B} T_{\rm g} / \mu}}, \label{eqn:condense}
\end{equation}
where $n_{\rm g} = \rho_{\rm g}/\mu$ and $\alpha$ is {the sticking probability}, i.e. fraction of dust-gas collisions that lead to growth. {Typical values for $\alpha$ are 0.1 \citep[see][and references therein]{vanLieshout2014}, which we use as our default}. The principle of detailed balance means that, at equilibrium, the rates of condensation and evaporation must be equal (e.g. \citealt{Langmuir1913}; {see also \citealt{Gail2013}}). Hence the evaporation rate is
\begin{equation}
    \left.\deriv{m}{t}\right|_{\rm evap} = 4 \upi a^2 \alpha \frac{P_{\rm v}(T_{\rm d})}{\sqrt{2\upi k_{\rm B} T_{\rm d} / \mu}}, \label{eqn:evap}
\end{equation}
where $P_{\rm v}(T_{\rm d})$ is the equilibrium vapor pressure. For the vapor pressure, we follow \citet{PB2013} and use parameters appropriate for forsterite, i.e. $P_{\rm v}(T) = \exp(-65308\,{\rm K}/T + 34.1)\,{\rm dyn}\,{\rm cm}^{-2}$.

The total dust formation rate {onto a population of dust grains with number density, $n_{\rm d}$,} is therefore
\begin{align}
    Q_{\rm d} &= n_{\rm d} \left(\left.\deriv{m}{t}\right|_{\rm cond}  - \left.\deriv{m}{t}\right|_{\rm evap} \right)  \nonumber \\
        & = \frac{3\rho_{\rm d}\alpha}{\rho_{\rm DUST} a} \left[\frac{P_{\rm g}}{\sqrt{2\upi k_{\rm B} T_{\rm g} / \mu}} - \frac{P_{\rm v}(T_{\rm d})}{\sqrt{2\upi k_{\rm B} T_{\rm d} / \mu}} \right], \label{eqn:dust_form}
\end{align}
with $Q_{\rm g} = - Q_{\rm d}$\footnote{We note that, in planetary atmospheres the growth rate maybe limited by the diffusion rate of the condensible species to the surface of the grains \citep[e.g.][]{Gao2018}. This is not the case in an atmosphere that is essentially made entirely of the condensible species, however.}. 

Although $Q_{\rm d}$ has been derived by considering the gain in mass of a dust grain due to condensation {applied to a population of grains}, we neglect any change in grain size and instead hold $a$ fixed. {This unusual application of \autoref{eqn:dust_form} requires further scrutiny, which we provide here. Our model should be thought of as a simple extension of an equilibrium condensation model, in which only the terms inside the square brackets are used to determine the impact of condensation, usually by removing the excess gas and placing it in the condensed form (or vice-versa) to maintain equilibrium. Equilibrium condensation is sometimes used in general circulation models \citep[e.g.][]{Ding2016}, or models of planet formation \citep[e.g.][]{Brouwers2018,Bodenheimer2018}. Similar models of condensation were applied to evaporating rocky planets by \citet{Castan2011,Nguyen2020,Nguyen2022} and \citet{Kang2021}. Our use of \autoref{eqn:dust_form} captures this equilibrium, while also allowing for deviations from equilibrium in regions where the timescale on which grains condense or evaporate is long compared to the flow timescale. Accounting for a finite sublimation time-scale is clearly necessary in our case because dust evaporation is thought to determine the length of the dusty tails seen by \emph{Kepler}/\emph{K2} \citep[e.g.][]{vanLieshout2014,Sanchis-Ojeda2015,vanLieshout2016}. Hence the outflows cannot be in condensation equilibrium at large distances from the planet. Since the growth and sublimation time-scale varies as $m/\dot{m} \propto a$, we must choose a value for the grain size. Here we use $1~\micron{}$ to be consistent with the observational constraints on grain size \citep[e.g.][]{Schlawin2021}. Assuming a large grain size means that our formation and destruction rates are slow and therefore our model will predict low dust abundances when growth is possible, but only to sub-\micron{} sizes. We discuss this further in \autoref{sec:fiducial} and \autoref{sec:composition}.}

{In applying \autoref{eqn:dust_form} as we do, we have completely avoided modelling the initial formation of dust grains, i.e. the nucleation of tiny seed particles or their injection into the outflow from the planet's surface. We simply assume that, should the conditions be right for macroscopic grains to be present as determined by \autoref{eqn:dust_form}, the formation of seed nuclei will proceed as necessary and the condensation / evaporation of dust from the surface of these grains will ultimately control the total mass of dust. We have however verified post-hoc that nucleation may plausibly occur in the models by considering the saturation ratio [i.e. $P_{\rm g}/P_{\rm V}(T_{\rm g})$]; saturation ratios substantially greater than unity (supersaturation) suggest nucleation may be possible.} 

The corresponding expressions for the momentum source term associated with dust formation, $Q'$, is given by momentum conservation,
\begin{align}
    Q'_{\rm d} = - Q'_{\rm g} = n_{\rm d}\left(u_{\rm g} \left.\deriv{m}{t}\right|_{\rm cond}  -  u_{\rm d} \left.\deriv{m}{t}\right|_{\rm evap}\right).
\end{align}
The energy sources include the exchange of thermal energy during condensation and evaporation, along with the release of latent heat
\begin{align}
    Q''_{\rm g} &= n_{\rm d} C_{\rm V, g} {\left(T_{\rm d} \left.\deriv{m}{t}\right|_{\rm evap} - T_{\rm g} \left.\deriv{m}{t}\right|_{\rm cond}\right)}, \\
    Q''_{\rm d} &= Q_{\rm d} L_{\rm s} + n_{\rm d} C_{\rm V, g} \left(T_{\rm g} \left.\deriv{m}{t}\right|_{\rm cond} - T_{\rm d} \left.\deriv{m}{t}\right|_{\rm evap} \right). 
\end{align}
where  $C_{\rm V, g}$ is the specific heat capacity of the gas and $L_{\rm s}=3.21\times 10^{10}\,{\rm erg\,g^{-1}}$ is the specific latent heat of sublimation. {Note that we assume the average energy of a gas particle created via evaporation of the dust grain is  $C_{\rm V, g}T_{\rm d}$, i.e. the new gas particles have the same temperature as the dust grain. This ensures detailed balance, as required for the evaporation/condensation process to eventually reach thermodynamic equilibrium (in the absence of other energy sources).}

\subsection{Heating and cooling}
\label{sec:heat_cool}
We include heating and cooling using the hybrid flux-limited diffusion approximation \citep{Kuiper2010}. In this approximation the radiative heating from the star is computed via plane-parallel ray-tracing while the thermal radiation emitted from the gas and dust is treated in the flux-limited diffusion approximation. Formally,
\begin{align}
    &\Gamma_s = \rho_s \kappa_{{\rm P}, s}(T_s) \left( 4\upi J - 4 \sigma T_s^4\right) +   \sum_{\nu_i} \rho_s \kappa_{{\nu_i},s} F_{{\nu_i},*} e^{-\tau_{\nu_i}}, \label{eqn:heating} \\ 
    &\tau_{\nu_i} = {\int_r^\infty} \sum_s \rho_s \kappa_{\nu_i,s} \, {\rm d}r, \\
    &\frac{4\upi}{c}\pderiv{J}{t} + \nabla \cdot F = - \sum_{s}\rho_s \kappa_{{\rm P},s}(T_s) \left[4\upi J - 4 \sigma T_s^4\right], \label{eqn:FLD} \\
    &F = - \frac{4\upi \lambda(J)} {\sum_s\rho_s \kappa_{{\rm R}, s}(T_s)} \nabla J, \label{eqn:FLD_flux}
\end{align}
where we have explicitly included a frequency dependence for the stellar irradiation, but used a grey model for the thermal radiation. This choice is motivated by the fact that the gas opacity contains strong lines that become optically thick high enough up in the outflow that they do not contribute significantly to driving the outflow, making a single mean opacity insufficient for the stellar heating. Grey FLD is however sufficient to capture the correct cooling rates due to thermal radiation. Here we denote the stellar flux in the band $\nu_i$ (where $\nu_i$ refers to the frequency) by $F_{{\nu_i},*}$, and the associated opacity of each species by $\kappa_{{\nu_i},s}$, with $\tau_{\nu_i}$ being the total optical depth at that wavelength. $J$ is the mean intensity of thermal radiation, and $\kappa_{{\rm P}, s}(T_s)$ and $\kappa_{{\rm R}, s}(T_s)$ are the Planck and Rosseland mean opacities of the species. $\lambda(J)$ is the flux-limiter, for which we follow \citet{Kley1989}, and $\sigma$ is the Stefan-Boltzmann constant.

For the dust opacity we adopt $\kappa_{\nu, {\rm d}} = 3Q(a, \nu)/4a\rho_{\rm DUST} $ where 
\begin{equation}
    Q(a, \nu) = 
    \begin{cases}
        1, & 2\upi a \nu \ge c, \\
        \left(2\upi a \nu/c\right)^{\beta}, & 2\upi a \nu < c,
    \end{cases} \label{eqn:dust_opacity}
\end{equation}
where we use $\beta = 1$ by default. For the corresponding Planck and Rosseland mean opacity we adopt the fitting function
\begin{equation}
    \kappa_{\rm P, d}(T) = \kappa_{\rm R, d}(T) =  \frac{3}{4a\rho_{\rm DUST}} \left[1 + \left({T_0}/{T}\right)^2\right]^{-\beta/2}
\end{equation}
where $T_0 = 458 (1 \micron / a) \,{\rm K}$ \citep{Owen2020}. The maximum error is $\sim 30$ per cent, but for dust temperatures above $T_0$ the error is $\lesssim 3$ per cent.

For the gas opacity, we assume a fixed composition based upon \citet{Schaefer2009}, as detailed in Appendix~\ref{app:Gas_Opac}. We include SiO, MgO, Fe, O, and Mg as opacity sources using the line-lists from ExoMol \citep{Li2019,Yurchenko2021} and \citet{Kurucz1992}. We have assumed local thermodynamic equilibrium for the level-populations and take into account natural and Doppler broadening, but neglect pressure broadening due to the low pressures considered here ($\lesssim 10^{-6}\,{\rm bar}$). 

For the stellar spectrum, $F_{\nu_i,*}$, we assume a Black-body with temperature $T_*= 4677$~K, appropriate for the Kepler-1520 system \citep[e.g.][]{Morton2016} and we use 41 variably-sized bins in the stellar irradiation calculation to capture the correct heating rate across the full range of column densities. See Appendix~\ref{app:Gas_Opac} for more details. 

\subsection{Boundary Conditions}
\label{sec:boundaries}

At the outer boundary we apply the standard outflow conditions for the gas, dust, and radiation, in which the flux is assumed constant across the boundary.

We take the inner boundary to be the planet's surface. For the planet's radius we use the mass-radius relation of \citet{Fortney2007}, assuming an earth-like composition (a rock mass fraction of 2/3). The inner boundary condition for the thermal radiation is that the outgoing thermal flux, $F_{\uparrow}$ is equal to the black-body emission form the planet's surface, $\sigma T_{\rm S}^4$, where $T_{\rm S}$ is the planet's surface temperature.  The details of how we compute  $F_{\uparrow}$ (and the thermal radiation received by the planet, $F_\downarrow$) in the flux-limited diffusion approximation are given in Appendix~\ref{app:FLD_updown}.

For the hydrodynamic equations, we use reflecting inner boundary conditions for $v_s$ and treat mass loss from the surface explicitly by adding mass and energy to the first cell as a source term. The mass-loss per-unit area from the planet's surface is given by
\begin{equation}
    \dot{\Sigma}  = 4 \alpha \left[\frac{P_{\rm v}(T_{\rm S})}{\sqrt{2\upi k_{\rm B} T_{\rm S} / \mu}} - \frac{P_{\rm g}}{\sqrt{2\upi k_{\rm B} T_{\rm g} / \mu}}\right] \equiv \dot{\Sigma}_{\rm evap} - \dot{\Sigma}_{\rm cond},
\end{equation}
where $T_{\rm S}$ is the surface temperature and $P_{\rm g}$, $T_{\rm g}$ refer to the pressure and density in the first active cell within the simulation and we identify $\dot{\Sigma}_{\rm evap}$, and $\dot{\Sigma}_{\rm cond}$ with the first and second terms, which correspond to evaporation and condensation at the surface. The rate of energy per unit area added to the cell due to condensation/evaporation is
\begin{equation}
   \dot{E} =  C_{\rm V, g}\left(T_{\rm S}  \dot{\Sigma}_{\rm evap} - T_{\rm g} \dot{\Sigma}_{\rm cond}\right).
\end{equation}

To compute the surface temperature, we model the energy change due to radiation, sublimation, and condensation via
\begin{align}
    C_{\rm S} \rho_{\rm DUST} \Delta r \pderiv{T_{\rm S}}{t} +  C_{\rm S} T_{\rm S} \dot{\Sigma} =& \sum_{\nu_i} F_{\nu_i, *} e^{-\tau_{\nu_i,{\rm S}}} + F_{\downarrow} - \sigma T_{\rm S}^4 \nonumber \\
    & + C_{\rm V, g}\left(T_{\rm g} \dot{\Sigma}_{\rm cond} - T_{\rm S}  \dot{\Sigma}_{\rm evap} \right) - L_{\rm s} \dot{\Sigma}, \label{eqn:boundary_T}
\end{align}
{where $C_{\rm S}$ is the specific heat capacity of the surface, which we take to be equal to the specific heat capacity of the dust.} Here the left hand side is the change in internal energy of the planet. The first three terms on the right hand side are the stellar and thermal radiation received by the planet, and its radiative cooling. The final terms are the thermal energy gained and lost due to condensation and evaporation, along with the latent heat.

To compute the change in the surface temperature, the thickness of the layer being heated and its heat capacity are needed. For a small body, the whole body will be isothermal at the same temperature as the surface, but this is not the case for a planet. Instead, we use the Biot number (which compares the heat flow through a body to the heat flow at its surface) to estimate the thickness of the surface layer, $\Delta r \sim \lambda_{\rm S} / \sigma T_{\rm S}^3 \approx 2.2\,{\rm cm} $ for a thermal conductivity, $\lambda_{\rm S} = 10^6\,{\rm erg\,s^{-1}\,cm^{-1}}\,{\rm K}^{-1}$ and $T=2000\,{\rm K}$ \citep[e.g.][]{Love1991,BookThermo, Dangelo2015}. This is clearly much smaller than the depth of any surface magma pool ($\sim 10\,{\rm km}$, \citealt{Kite2016}), but the turnover time of the magma is also long, thus  transport within the magma does not significantly perturb the surface temperature \citep{Kite2016}, which ultimately reaches equilibrium. Since the precise value has no impact on the steady-state mass-loss rates (because $\pderiv{T_{\rm S}}{t} = 0$), we simply use $\Delta r = 10\,{\rm cm}$. For this choice, the surface temperature reaches equilibrium with the radiation on a time-scale of $\sim 10^3\,{\rm s}$.

\subsection{Numerical Method}

To solve these equations, we use the multi-species finite-volume code \textsc{aiolos} \citep{Schulik2022}. \textsc{aiolos} solves the equations of gas dynamics, including gravity, using the well-balanced Godunov method of \citet{Kappeli2016} and an HLLC Riemann solver \citep[e.g.][]{Toro}. For the dust species, we skip the well-balancing step (because the dust is never close to hydrostatic equilibrium) and use the Riemann solver of \citet{Pelanti2006}. The gradients are reconstructed at second order with the slopes limited using the MC-limiter of \citet{Mignone2014} and time integration is done using the second order strongly stability preserving Runge-Kutta method of \citet{Gottlieb}. The dust-gas coupling terms are solved using the semi-implicit method outlined in \citet{Benitez-Llambay2019} and radiation transport is solved using the linearization method of \citet{Commercon2011}, with the stellar irradiation source terms treated as in \citet{Bitsch2013}. The source terms, $Q_s, Q'_s,$ and $Q''_s$ are evaluated implicitly using an operator-split approach, as described in Appendix~\ref{app:numerics}

{We initialize the simulation with a non-zero dust density ($10^{-20}\rho_g$) because the growth rate in \autoref{eqn:dust_form} is zero when the dust density is zero. For the initial gas density, temperature, and velocity we have assumed hydrostatic equilibrium at the optically thin dust temperature (see below). Our results presented are not sensitive to these choices: in models with efficient dust formation we will always find that the gas and dust reach condensation-evaporation equilibrium close to the planet, thus losing any memory of their initial condition. Further, the condition for growth ($Q_{\rm d} > 0$) is not sensitive to the amount of dust present when the dust is optically thin, which is the case for our initial conditions.}

\begin{figure*}
	\includegraphics[width=\textwidth]{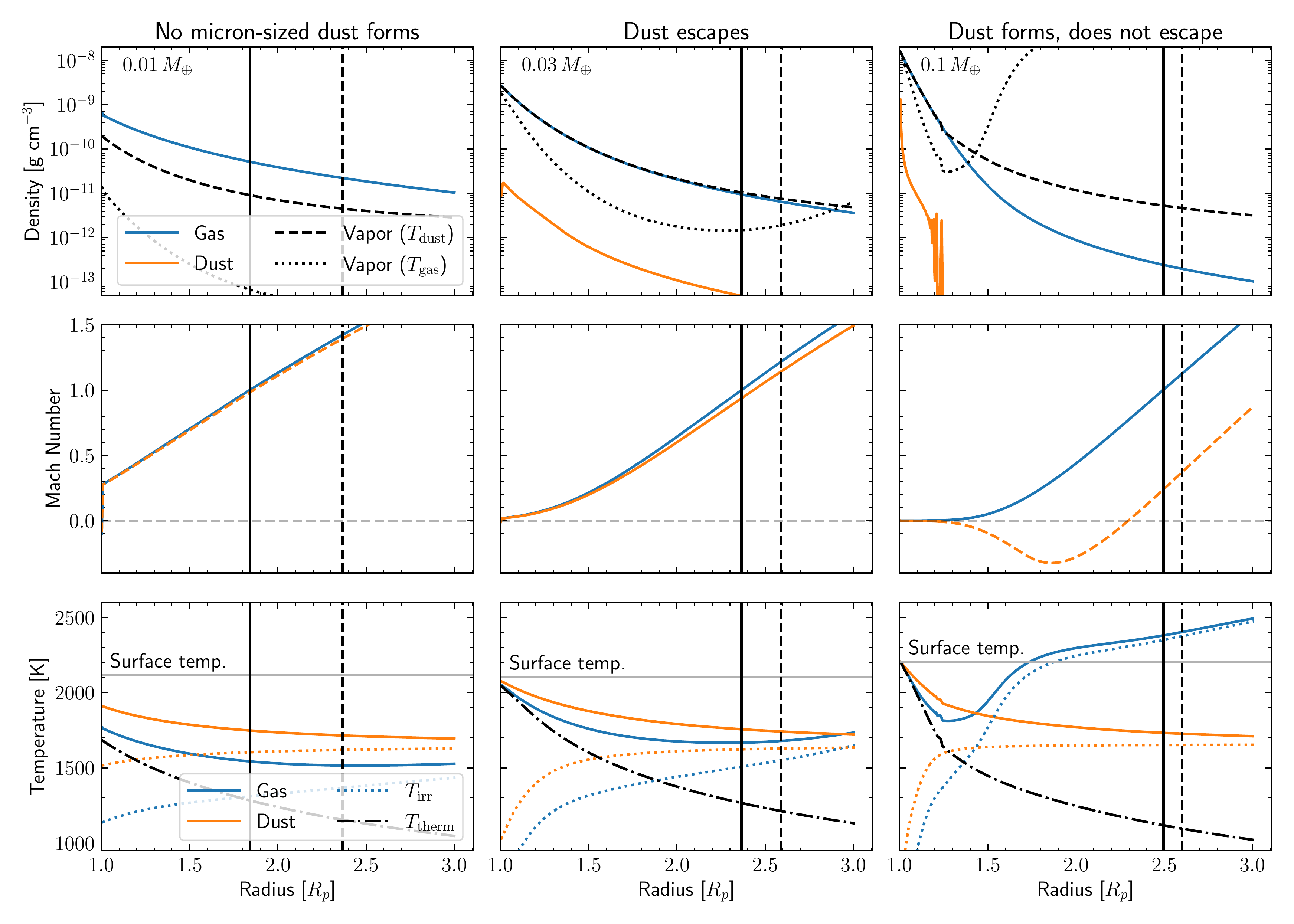}
    \caption{Wind structure for three different planet masses for the fiducial parameters. {\bf Top}: the gas and dust density. The density that would be obtained at vapor pressure equilibrium for dust at the gas temperature {[$\rho_{\rm vapor}(T_{\rm g}, T_{\rm g})$] or dust temperature  [$\rho_{\rm vapor}(T_{\rm g}, T_{\rm d})$] is also shown}. {\bf Middle}: the Mach number in the flow. {The Mach number of the dust is shown as a dashed line when the dust-to-gas ratio is below $10^{-5}$.} {\bf Bottom}: The temperature of the gas and dust. The horizontal grey lines show the planet's surface temperature. For comparison, the temperatures that the dust and gas would obtain if only heating due to the attenuated stellar irradiation (dotted) or re-emitted thermal radiation (dot-dashed) was included are also shown. The vertical solid and dashed lines denote the sonic point and Hill radius, respectively. {Gas mass-loss rates measured at the Hill radius are 3, 2 and $0.2\,M_\oplus\,{\rm Gyr}^{-1}$ for the 0.01, 0.03, and $0.1\,M_\oplus$ planets respectively. The dust mass-loss rate is $0.01\,M_\oplus\,{\rm Gyr}^{-1}$ for the $0.1\,M_\oplus$ planet and negligible for the others.}}
    \label{fig:flow_fiducial}
\end{figure*}

\subsection{Summary and comparison to previous models}

We use a 1D multi-species radiation-hydrodynamic model for the dusty outflows. Our model includes radiative heating and cooling of both the gas and dust, including both the stellar radiation and the re-emitted thermal radiation (treated via FLD) of the dust and gas. We also include a physical description of dust formation and destruction, including latent heat release during dust formation. We also explicitly include a model for the planet's surface.

Compared to \citet{PB2013}, our model differs in that we include radiative heating of both the gas and dust, including both heating from the stellar irradiation and the re-emitted thermal radiation, while \citet{PB2013} included only stellar heating on the dust -- leading to dust temperatures that decrease towards the planet. They also parameterized the formation of dust, while we include a minimal physical model. 

Recently \citet{Kang2021} developed a pseudo-2D model that includes flows from the day side to the night side. Their mass-loss model from the day side is however a 1D model, similar to ours. We include a similar treatment of dust formation, except that \citet{Kang2021} assume equilibrium for the dust abundance, while we parameterize the growth/destruction rate. \citet{Kang2021} did not explicitly model the temperature structure, but parameterized it. {By including both radiative transfer and a model for dust formation, we are able to discrimanate between the conclusions of \citet{PB2013} (who suggested dust will be present because it may nucleate) and \citet{Kang2021} (who suggested that dust would not survive on the day sides due to high dust temperatures).} 

\section{Results}
\label{sec:results} 

\begin{figure}
	\includegraphics[width=\columnwidth]{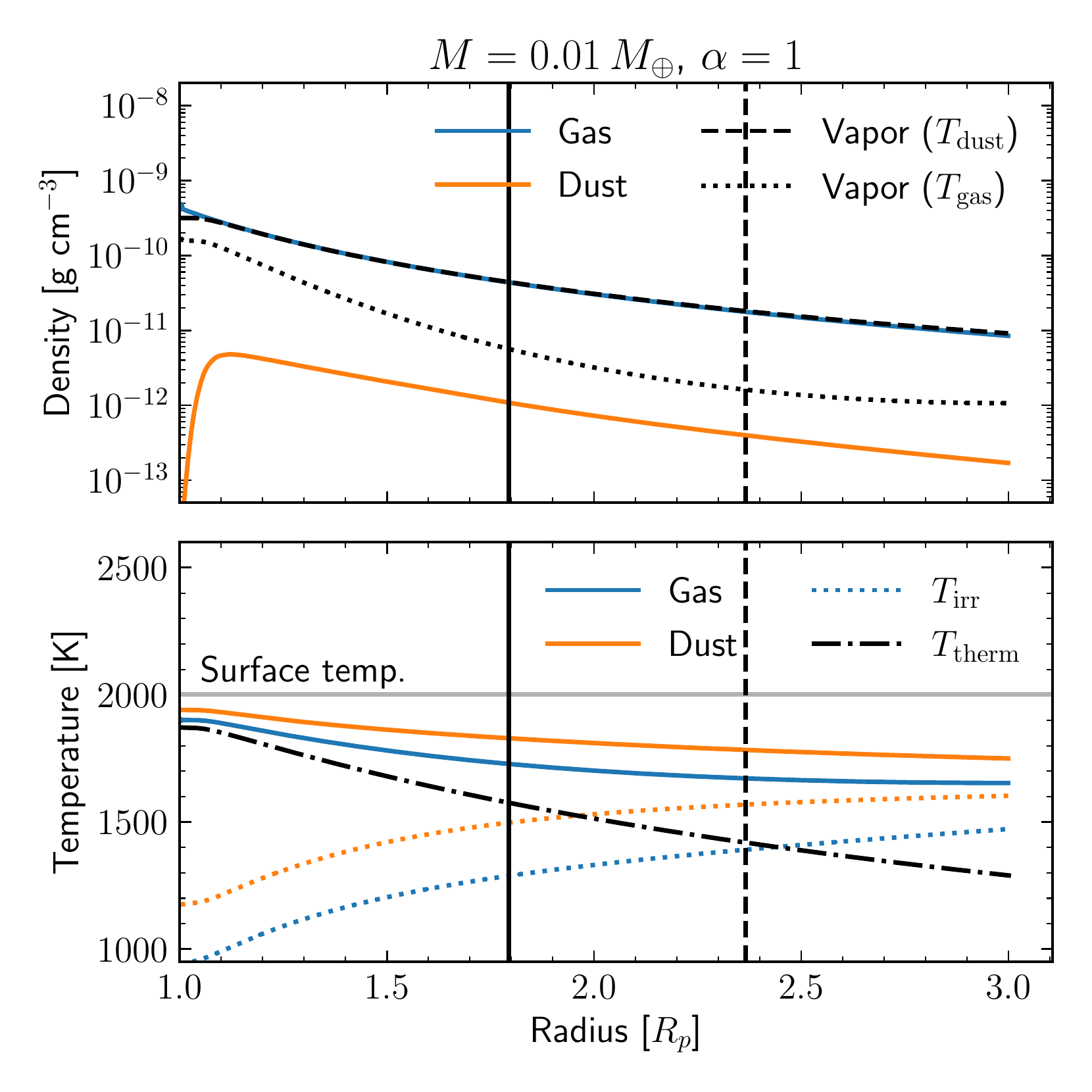}
    \caption{Same as \autoref{fig:flow_fiducial}, except the dust growth rate was increased by setting the fraction of collisions that lead to growth to unity ($\alpha=1$) instead of the fiducial value of 0.1. Dust forms readily given the higher growth rate, {with a mass-loss rate of $0.05\,M_\oplus\,{\rm Gyr}^{-1}$.}} 
    \label{fig:high_alpha}
\end{figure}

\subsection{Fiducial model: favourable conditions for dust formation}
\label{sec:fiducial}

We begin by examining the outflows in our fiducial model, for which we have chosen parameters appropriate for a K-type star like Kepler-1520. We assume a stellar mass, luminosity, and effective temperature of $M_* = 0.76\,M_\odot$, $L_* = 0.22\,L_\odot$, and $T_* = 4677\,{\rm K}$, respectively \citep[e.g.][]{Morton2016}. The planets are placed on a circular orbit at 0.0134~au. We assume that the planet is tidally locked and heat redistribution is inefficient, electing to apply our model at the sub-stellar point. The dust grain size is taken to be $1\,\micron$, as suggested by observations \citep[e.g][]{Brogi2012,Budaj2013,Sanchis-Ojeda2015,Schlawin2021}. The fiducial model for three different planet masses (0.01, 0.03, and $0.1\,M_\oplus$) are shown in \autoref{fig:flow_fiducial}.

{\autoref{fig:flow_fiducial} shows three different behaviours for the different planet masses. For the lowest mass planet, a negligible amount of dust has formed. Dust forms close to the other two planets, but only in the intermediate mass case is there still a significant amount of dust once the outflow reaches the planet's Hill radius. To explain these behaviours it is useful to start by considering the lowest mass planet first.}

{In the top panels of \autoref{fig:flow_fiducial} we plot the gas and dust densities, along with two other curves that show the expected gas density if it were in condensation-evaporation equilibrium (\autoref{eqn:dust_form}). Explicitly, the condition for equilibrium is $P_{\rm g}/\sqrt{T_{\rm g}} = P_{\rm v}(T_{\rm d}) / \sqrt{T_{\rm d}}$, from which we find}
\begin{equation}
    \rho_{\rm vapor}(T_{\rm g}, T_{\rm d}) = \frac{\mu}{k_{\rm B}} \frac{P_{\rm v}(T_{\rm d})}{\sqrt{T_{\rm g}T_{\rm d}}}.
\end{equation}

{Comparing the gas density and $\rho_{\rm vapor}(T_{\rm g}, T_{\rm d})$ (the dashed lines) for the lowest mass planet, we see that the gas density exceeds $\rho_{\rm vapor}(T_{\rm g}, T_{\rm d})$ and therefore we have $Q_{\rm d} > 0$. The conditions are thus favourable for dust formation in the sense that, if left for a sufficient length of time at those temperatures and pressures, the gas should condense and form dust. The reason that a substantial amount of dust does not form in this case is simply that the time required for this to happen is longer than the time that the gas spends under such conditions, i.e. the growth time-scale is longer than the local flow time-scale. This can be verified simply by considering a model with a shorter growth time-scale. One such model is shown in \autoref{fig:high_alpha}, where the growth time was decreased by increasing the $\alpha$-parameter in \autoref{eqn:dust_form}. In this case we see that dust forms readily. Since the growth time-scale depends on the grain size, these results can also be interpreted as saying that, while the grains are growing in the fast outflows from low mass planets, they do not reach the $1\,\micron{}$ size assumed in the model.  We confirm this by directly integrating the equations that govern the growth of a single dust grain (\autoref{eqn:condense} and \autoref{eqn:evap}) along the flow. From this we find the grains only reach a size of $0.2\,\micron{}$.}

{Examining further the intermediate mass (0.03\,$M_\oplus$) case, we see that dust has formed readily. We also see that $\rho_g \approx \rho_{\rm vapor}(T_{\rm g}, T_{\rm d})$, indicating $Q_{\rm d} \approx 0$, i.e. condensation has occurred sufficiently fast that the dust and gas densities have reached condensation-evaporation equilibrium. By the time that the flow has reached the Hill radius the gas density has dropped enough that $\rho_g < \rho_{\rm vapor}(T_{\rm g}, T_{\rm d})$ and thus the destruction of the dust grains has begun. However, the outflow speeds are sufficiently high that the dust escapes before a significant fraction of it has been destroyed.}

{Turning finally to the highest mass case shown in \autoref{fig:flow_fiducial} the behaviour is essentially the same as the intermediate mass case. The key exception is that the transition to dust destruction occurs closer to the planet where the outflow velocities are slow enough that the dust is destroyed before it can escape. The middle-right panel of \autoref{fig:flow_fiducial} shows that this is inevitable for the 1~$\micron{}$-sized grains used here because they are falling towards the planet at the point where the transition to destruction occurs.}

{Smaller grains are more easily dragged outwards by the flow since they are more tightly coupled to the gas; 0.1~\micron{} grains have outward directed velocities everywhere (not shown), even for the  highest planet mass in \autoref{fig:flow_fiducial}. Despite this, the wind would still be free of sub-\micron{} grains when it escapes the Hill radius of the highest planet mass because the outflow velocities are slow and therefore the grains have sufficient time to evaporate.}

{Next, we consider whether the seed nuclei, onto which dust condenses, can form under the conditions present in the wind. \autoref{eqn:dust_form} does not address this directly, since it is derived based on the growth and destruction of pre-existing grains. A rough way to determine whether nucleation is likely to occur is to look at the saturation ratio, $S=P_{\rm g}/P_{\rm v}(T_{\rm g})$. Supersaturation, $S > 1$, suggests that the initial step in dust formation, nucleation (neglected in our model), should be possible. We demonstrate that our models do indeed predict supersaturation by plotting $\rho_{\rm vapor}(T_{\rm g}, T_{\rm g})$ as dotted lines in \autoref{fig:flow_fiducial}; the ratio of the $\rho_{\rm g} / \rho_{\rm vapor}(T_{\rm g}, T_{\rm g})$ is equivalent to $S$.  Close to the planet $\rho_{\rm vapor}(T_{\rm g}, T_{\rm g})$ is typically below the gas density, demonstrating that the gas is supersaturated, suggesting that nucleation may occur (as was also suggested by \citealt{PB2013}). Far away from the planet the gas temperature can exceed the dust's temperature and we see that the saturation ratio suggests that nucleation is no longer favoured. However, we acknowledge that nucleation is sensitive to the conditions present, and therefore should be modelled in detail in the future.}

{Next we explore why our models produce conditions that result in dust growth close to the planet. The explanation is simply that the dust's temperature in the wind} is lower than the planet's surface temperature. Since the wind is produced by evaporating the planet's surface, we have $P_{\rm g}\approx P_{\rm v}(T_{\rm S})$ at the surface and the evaporation and condensation rate from the planet's surface are approximately balanced. However, since the dust temperature is lower than the planet's surface temperature, the evaporation rate from the surface of dust grains is lower the condensation rate, leading to dust formation\footnote{The evaporation and condensation rates per unit area would be the same for the dust grains and the planet, if they had the same temperature.}. This can be seen immediately from the left panels of \autoref{fig:flow_fiducial}, where the gas density (blue lines) is higher than the density that would be achieved if it were in equilibrium with the vapor pressure from dust evaporation (black dashed lines).

The origin of the low dust temperatures compared to the planet's surface temperature can be explained by considering the main sources of heating and cooling for the planet and the dust grains. Far from the planet, the dust temperature is set by the balance between equilibrium between stellar irradiation and cooling, i.e. $\kappa_{\rm P, d}(T_*) \sigma T_*^4(R_*/a_{\rm p})^2 = 4 \kappa_{\rm P, d}(T_{\rm d}) \sigma T_{\rm d}^4$. For 1~\micron{} grains, $\kappa_{\rm P, d}(T_*) \approx \kappa_{\rm P, d}(T_d)$, and therefore $T_{\rm d} \approx (R_*/2a_{\rm p})^{1/2} T_* \approx 1650\,{\rm K}$ {(sub-\micron{} grains, whose opacity is non-grey, will deviate from this estimate; they are considered in detail in the next section)}. Close to the planet the dust is warmer than this by a factor of $\sim 2^{1/4}$ because of the additional heating from the planet's emitted radiation\footnote{This is opposite to what \citet{PB2013} assumed; they assumed that the dust would be cooler closer to the planet.}.  The planet is however even warmer than the dust because heat redistribution inside the planet is inefficient. Thus the planet's surface only cools through $\upi$ steradians  resulting in a temperature at the sub-stellar point that is a factor of $4^{1/4}$ warmer than the dust temperature at large distances from the planet (compare \autoref{eqn:heating} with \autoref{eqn:boundary_T}). In practise the planetary surface temperatures in the simulations are slightly lower than this, because some of the stellar radiation has been absorbed by the intervening gas and dust. {A full radiation-hydrodynamic model, such as the one that we have presented, is needed to properly account for these complications.}

\begin{figure}
	\includegraphics[width=\columnwidth]{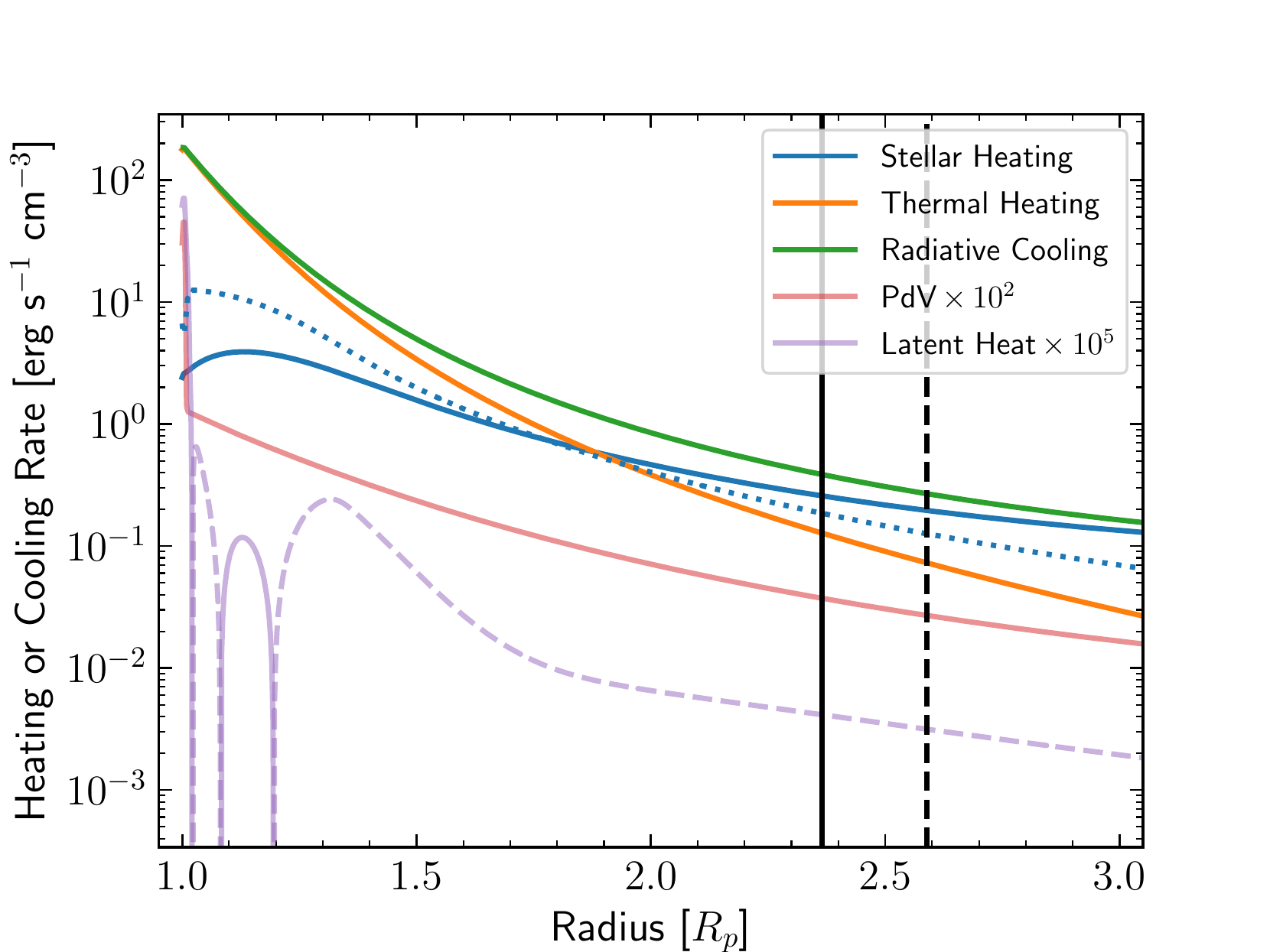}
    \caption{{The various heating and cooling rates of the gas due to different processes for the $0.03\,M_\oplus$ fiducial model. Radiative heating (due to stellar irradiation and re-absorbed thermal radiation) and cooling processes dominate over $P{\rm d}V$ work. The blue dotted line shows the energy input through the stellar radiation heating the dust. We also show the heating and cooling due to latent heat associated with dust formation, with the dashed lines referring to regions where dust is being destroyed, resulting in a net cooling. Note that because the latent heat release initially heats the dust grains, this energy is re-radiated rather than directly heating the gas.}}
    \label{fig:heating} 
\end{figure}

The gas temperature in the outflow is similarly controlled by the balance between radiative heating and cooling, {as shown in \autoref{fig:heating}. Close to the planet, the gas heating is dominated by the absorption of thermal radiation. Cooling due to $P{\rm d}V$ work is negligible. The latent heat released during dust condensation, which initially heats the dust grains, can indirectly heat the gas since it contributes to the thermal radiation; however, our results indicate this is also negligible. Friction and the collisional heat exchange (\autoref{eqn:coll_en}) between the dust and the gas is even smaller than the $P{\rm d}V$ work (not shown). The dust does indirectly contribute a significant heating of the gas, which happens predominately through the re-emission of the stellar radiation that it absorbs.} Far from the planet, the gas reaches high temperatures ($\sim 3000\,{\rm K}$) due to heating from the strong iron lines in the optical. The stellar radiation in these lines is quickly extincted though, leading to lower stellar heating rates closer to the planet's surface {because only regions of the spectrum for which the gas opacity is lower still contain significant amounts of flux. This causes the heating of the gas by stellar irradiation to fall faster than the heating of the dust, for which the opacity is close to grey (see also Appendix~\ref{app:Gas_Opac}). Finally, since the heating due to thermal radiation emitted by the dust and the planet decreases outwards, the combination of stellar and thermal heating typically results in the gas temperature having a minimum close to the location where the stellar irradiaion and thermal heating are equally important.}

{It is possible that the gas in these outflows could be detected via the transit method using these strong absorption lines since they become optically thick at several planetary radii. We refrain from predicting transit depths since ionization is likely to be important at such altitudes \citep[e.g.][]{Ito2021}, and note that gas absorption lines have not yet been detected in systems with dusty tails \citep[e.g.][]{Ridden-Harper2019}.}

\subsection{Small grains, realistic opacities, and heterogeneous condensation}
\label{sec:composition}

\begin{figure}
	\includegraphics[width=\columnwidth]{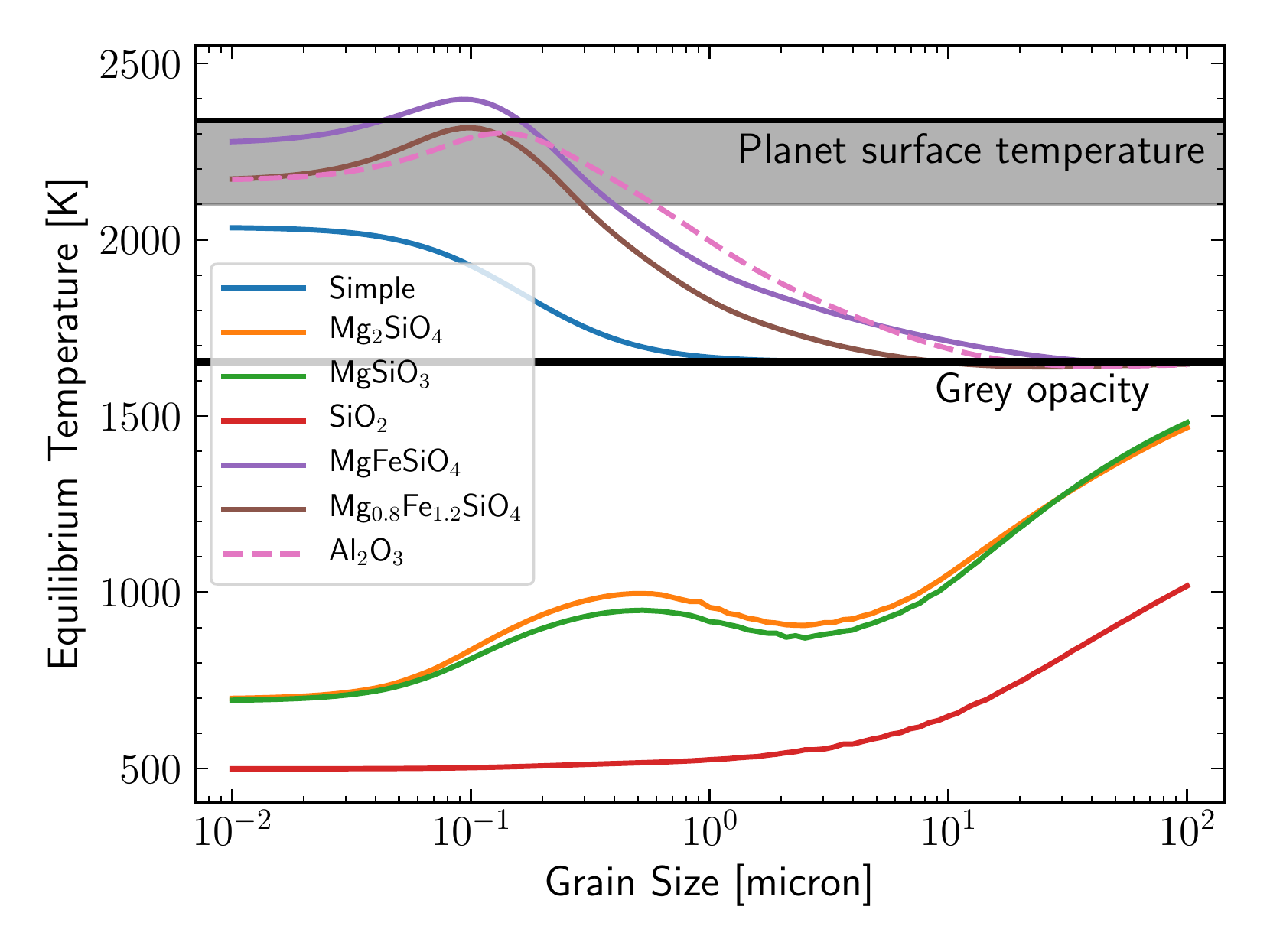}
    \caption{The dust temperature in the optically thin limit as a function of grain size for the simple opacity model (\autoref{eqn:dust_opacity}) and various real condensates that might be expected to from in dusty winds. The black lines show the planet temperature at the sub-stellar point and the temperature of dust with a grey opacity. Due to modest optical depths, the temperatures in the full models are 100 to 200~K lower (grey band). At sizes $\gtrsim 1\,\micron{}$ the dust temperatures are below the planet surface temperature favouring dust formation, but for smaller sizes only iron-free silicates have sufficiently low temperatures.} 
    \label{fig:Tdust}
\end{figure}

We have shown that dust formation occurs when dust temperatures are lower than the planet's surface temperature and demonstrated that grey dust opacities provide the necessary conditions. Small grains however do not provide a grey opacity, resulting in an optically thin dust temperature of 
\begin{equation}
    T_{\rm d} \approx \left[\frac{\kappa_{\rm P, d}(T_*)}{\kappa_{\rm P, d}(T_{\rm d})} \frac{R_*^2}{4a_{\rm p}^2}\right]^{1/4} T_* \approx 1650\left[\frac{\kappa_{\rm P, d}(T_*)}{\kappa_{\rm P, d}(T_{\rm d})}\right]^{1/4}\, {\rm K}.
\end{equation}
The temperature of small grains will therefore be hotter than the temperature of large grains when the dust opacity at optical wavelengths, where the stellar irradiation peaks, is larger than the opacity at near-infrared wavelengths, through which the dust cools. This is the case for our simple opacity model, and thus prevents dust from forming in the models {when we use grain sizes $\lesssim 0.1\,\micron{}$}. We next consider whether this is the case for realistic opacities associated with reasonable condensates.

\autoref{fig:Tdust} shows the optically thin equilibrium temperature of dust grains with different grain sizes for both our simple opacity model and a range of condensates. We computed the opacity for spherical grains with realistic compositions using the \citet{Mie} Mie-theory code in the \textsc{dsharp\_opac} library \citep{Birnstiel2018}. The refractive indices were taken from \citet{Kitzmann2018}. 

For sufficiently large grains, the opacities become grey leading to the same equilibrium temperature. For small grains there is a clear dichotomy. Iron-free silicates (MgSiO$_3$, Mg$_2$SiO$_4$, SiO$_2$) are all transparent at optical wavelengths but have a high opacity at 10~\micron{} due to the silicate feature, leading to low equilibrium temperatures. Iron-rich silicates have more opacity in the optical than the near-infrared leading to higher temperatures for the small grains, thus behaving more similarly to the simple opacity model \citep[see also][]{Ossenkopf1992}. We also show corundum (Al$_2$O$_3$). Corundum is not expected to be present in large quantities because little aluminium is present \citep[e.g.][]{Schaefer2009}, but previous modelling of the dusty-tail properties favoured a corundum composition over silicates \citep{vanLieshout2014}. It is also clear that corundum formation is disfavoured too, due to the high temperatures of small corundum grains.
 
The dust temperature results for different compositions favour a specific scenario for dust condensation in these winds, which we use our results to sketch out. Considering first small grains, the iron-rich silicates will reach high temperatures and evaporate, while grains made of predominantly iron-free  silicates will have low temperatures and therefore be stable. Forsterite (Mg$_2$SiO$_4$) and silica (SiO$_2$) have the lowest vapor pressures of the materials considered, making them the most stable condensates (e.g. \citealt{vanLieshout2014}). Thus, magnesium-rich silicates will likely condense before iron or iron-rich silicates. These grains will initially be small, and will obtain low temperatures due to their low iron content. Their low temperatures will allow heterogeneous condensation and iron may be incorporated into the grains. {This picture of an initially low iron abundances as the grains start to form is consistent with models and observations of dust formation in AGB star winds or protoplanetary discs \citep{Gail1998,Molster2002,Lodders2003,Gail2013,Blommaert2014}. Due to the intense stellar radiation present in the case studied here, the higher optical opacity of iron-rich grains would cause the grains' temperature to rise as the iron abundance increases.} However, because iron evaporates from silicates more readily than magnesium \citep[e.g.][]{Costa2017}, increasing the dust temperature will cause the iron content to decrease.  Therefore, the small grains may reach an equilibrium composition controlled by the feedback between iron content and dust's temperature. Although low growth rates could prevent an equilibrium being reached {(which would result in grains with a lower iron abundance)}, the growth time is shorter than the flow time-scales for grain sizes $\lesssim0.1\,\micron$. Finally, as the grains grow to larger sizes the opacity will become greyer and the temperature will decouple from the composition. In this way, we expect that grain condensation will proceed without difficulty, ultimately producing the observed micron-sized grains. 

{During heterogenous condensation the gas composition will also change: the removal of MgO and SiO would cause the gas opacity in the infra-red to drop and the gas temperature to increase, while the removal of iron will remove optical opacity causing the gas to cool. We may expect significant changes in the gas temperature to be associated with heterogeneous condensation, with the formation of iron-poor silicates leading to higher gas temperatures.}

The question still remains as to whether more massive planets ($\gtrsim 0.1\,M_\oplus$) are able to produce dusty outflows. Our models for these planets certainly predict that dust grains should form close to the planet, but not at larger distances and micron-sized grains were not entrained in the wind. Heterogeneous condensation and evaporation may avoid the problem of grain destruction since pure forsterite or silica would attain low enough temperatures that they will survive, but a detailed model for the grain sizes is needed to determine whether they have appropriate sizes to be entrained. Nevertheless, it is possible that any dust escaping from the day-side of $\sim 0.1\,M_\oplus$ planet would have a low optical opacity and a higher infrared opacity to obtain the required low temperatures. {This would not necessarily help explain the observed tails, which are thought to have grey opacity \citep[e.g.][]{Schlawin2021}}. 

\subsection{Sensitivity to planet surface temperature}
\label{sec:grid}

\begin{figure*}
	\includegraphics[width=\textwidth]{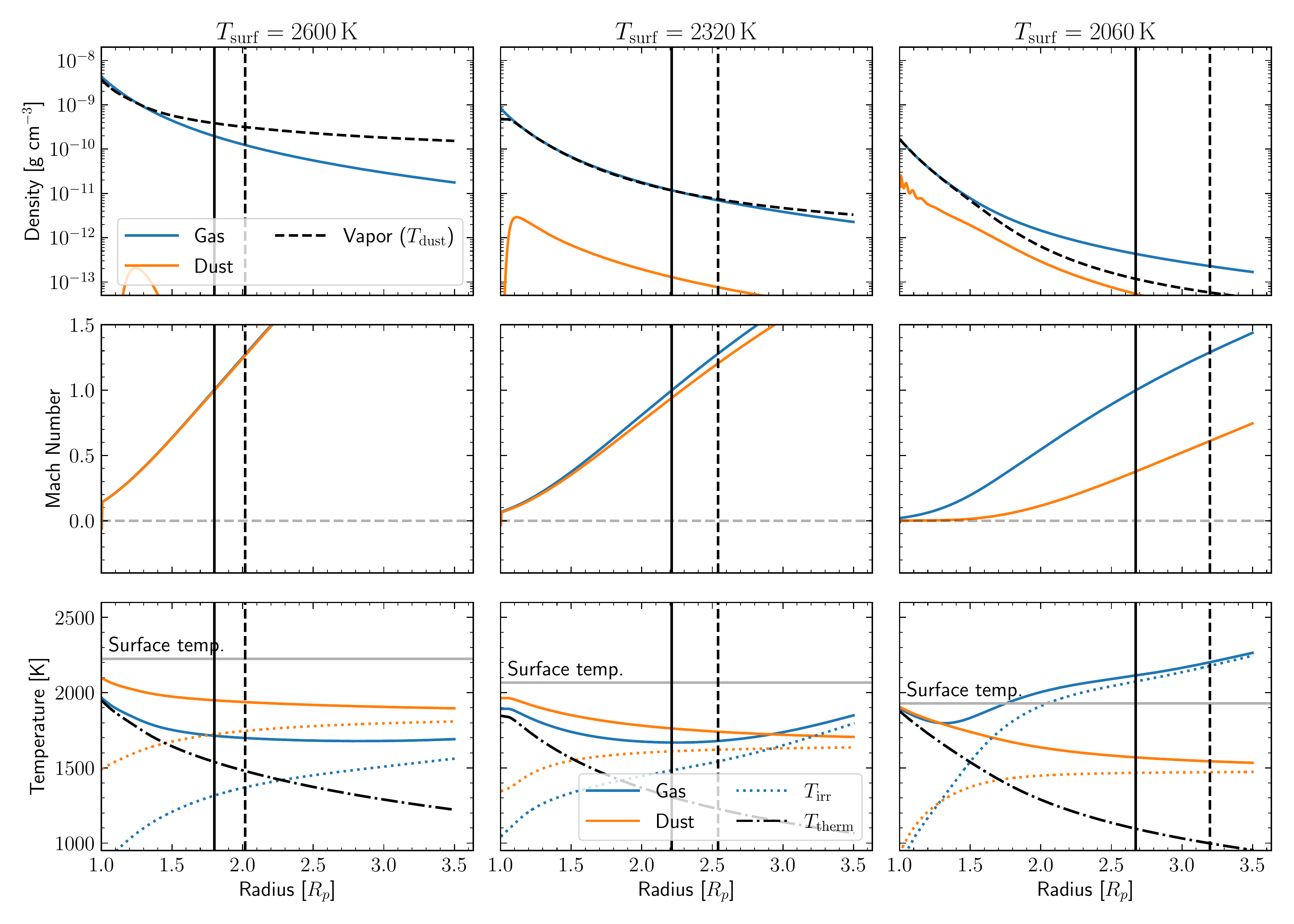}
    \caption{Like \autoref{fig:flow_fiducial}, except for $0.02\,M_\oplus$ planets with different surface temperatures. {Note that the label $T_{\rm surf}$ is an estimate of the surface temperature assuming the outflow is optically thin and has a Bond Albedo of zero. I.e. it is $4^{1/4}$ times hotter than the planet's equilibrium temperature. In practice the planet's surface temperatures are somewhat cooler. Gas mass-loss rates measured at the Hill radius are 10, 1 and $0.1\,M_\oplus\,{\rm Gyr}$ for the $T_{\rm surf}=2600$, 2320, and $2060\,{\rm K}$ cases, respectively. The dust mass-loss rates are 0.01 and $0.006\,M_\oplus\,{\rm Gyr}^{-1}$ for the 2320, and 2060~K cases and negligible for the 2600~K planet.}} 
    \label{fig:T_compare}
\end{figure*}

\begin{figure}
	\includegraphics[width=\columnwidth]{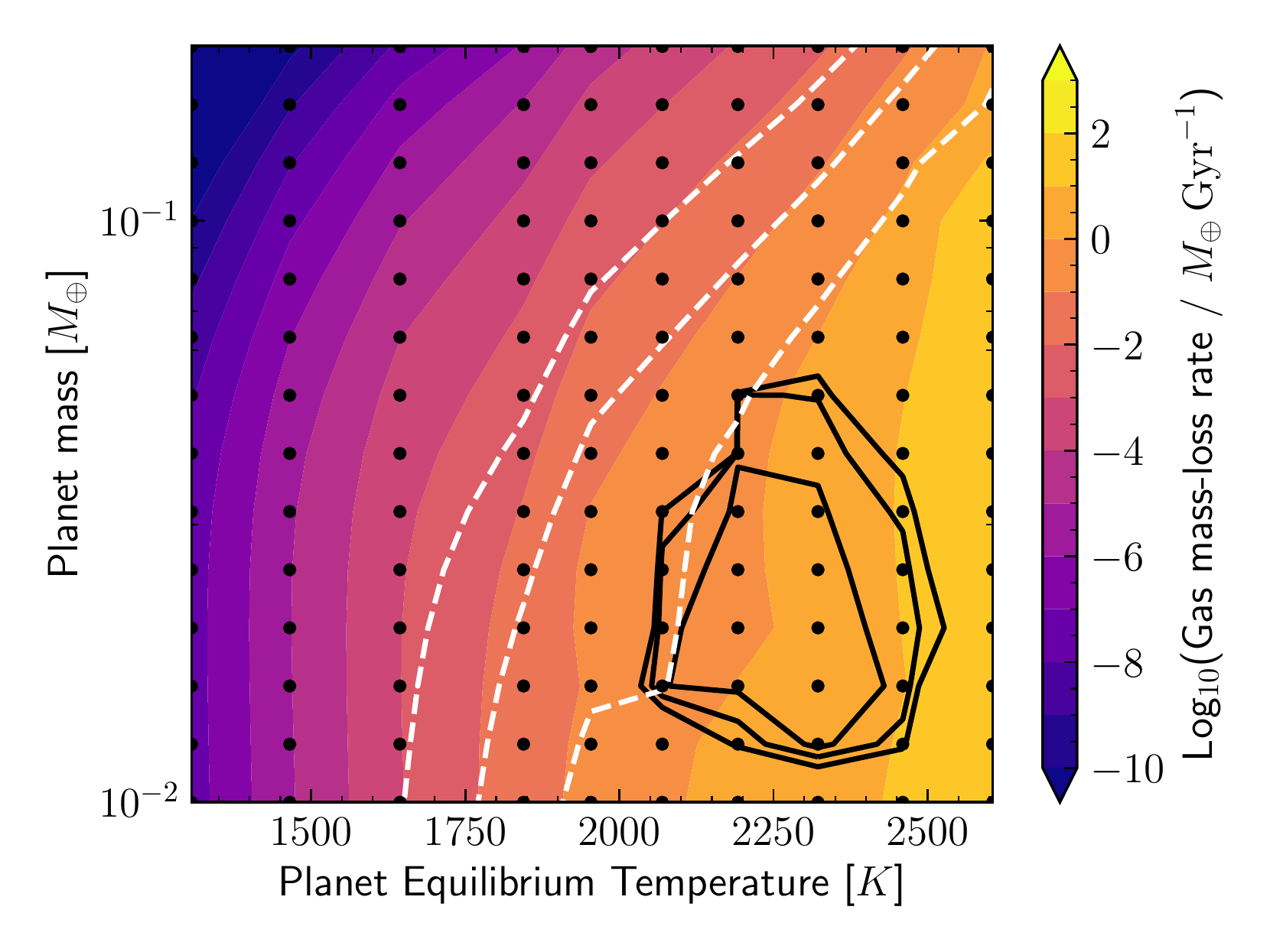}
    \caption{Gas mass-loss rates for different planet masses and equilibrium temperatures. The black points show the individual simulations and the contours denote the region where the outflow contains a significant amount of dust, i.e., dust mass-loss rates exceeding $10^{-4}$, $10^{-3}$, and $10^{-2}\,M_\oplus\,{\rm Gyr}^{-1}.$ These are below estimated of the observed mass-loss rate, see \autoref{sec:grid}. The white dashed lines show locations where the mass-loss timescale $M_{\rm p}/\dot{M}_{\rm p}=0.1$, 1 and 10~Gyr.} 
    \label{fig:mass_loss}
\end{figure}

In the previous sections we have explored models for planets orbiting a star with $M_*=0.76M_\odot, L_* = 0.22\,L_\odot$, $T_* = 4677\,{\rm K}$ at $a = 0.0134$~au, which corresponds to an effective temperature of $T_{\rm surf}=T_*(R_*/a)^{0.5} = 2320\,{\rm K}(a/0.134\,{\rm au})^{-0.5}$ at the sub-stellar point\footnote{As discussed before the temperature that the planet's surface actually obtains will generally be lower than this due to the non-negligible optical depth in the outflow.}. Here we vary the semi-major axis of the planet's orbit to explore the impact of the surface temperature. It should be noted that changing the semi-major axis also affects the planet's Hill radius, thereby affecting the mass-loss rate because the Hill radius is typically small enough to affect the outflow's sonic point \citep[see, e.g.,][]{PB2013}. In other words, the mass-loss rates at a given planet temperature will depend on the star's spectral type, because the mass-loss rates are sensitive to both the stellar temperature and mass (i.e. tidal gravity).

Different surface temperatures primarily affect the gas mass-loss rates of gas through the change in pressure at the surface (\autoref{fig:T_compare}), which is set by vapor pressure equilibrium. Hotter planets therefore have more dense outflows. Higher temperatures also produce faster outflows, further increasing the mass loss rates. 

Perhaps surprisingly, we find that the {hottest planets (i.e. $T_{\rm surf} \gtrsim 2500\,{\rm K})$} have less dusty outflows. {In \autoref{fig:T_compare} the difference between the fiducial and hotter cases (middle and left rows of \autoref{fig:T_compare}) is driven by two competing effects: the higher temperatures resulting in faster outflows and the higher gas pressures producing lower grain growth time-scales. Of these two effects the faster outflow is more significant, resulting in less dust formation. For more massive planets (not shown), the same behaviour as in the fiducial case is found, i.e. the outflow is slower for more massive planets resulting in more dust formation closer to the planet. } The two cooler planets shown in \autoref{fig:T_compare} have similar dust mass-loss rates despite the density difference. The higher dust density in the cooler simulation is therefore driven primarily by the lower outflow speed rather than different dust production rates. Further, dust production at $T_{\rm surf}= 2000\,{\rm K}$ is limited by the long growth time-scales associated with the low densities, with the conditions being favourable for growth far beyond the planet's Hill sphere. At even lower temperatures the growth time-scales become too long and the outflows are too weak to entrain micron sizes grains, resulting in dust free winds. 

In \autoref{fig:mass_loss} we show the mass-loss rates for the gas {for different planet masses and equilibrium temperatures ($T_{\rm surf}$). We} denote by contours the region in which our model is able to produce a significant quantity of 1~\micron{} grains. When computing the mass-loss rates we have assumed that the winds originate from the day side only, covering $\upi$ steradians. 

At low planet masses, $\lesssim 0.05M_\oplus$, the gas mass-loss rates are not sensitive to the planet's mass but primarily depend on temperature through its influence on the pressure at the surface. Increasing the planet's mass further causes a reduction in the mass-loss rate because the Bondi radius and Hill radius increases, as expected for a Parker wind model. Similar results were seen by \citet{PB2013}.

The formation of 1~\micron{} sized-dust in the wind is limited by a few factors. At low planet masses, $\lesssim 0.01\,M_\oplus$, the primary limit is that the growth time of micron-sized grains is longer than the outflow time scale. Sub-micron grains may be able to form, however, extending the region of parameter space over which the outflows are dusty to lower temperatures and planet masses. For masses $\gtrsim 0.05M_\oplus$ dust formation is instead limited by the steep pressure gradient in the outflow (see \autoref{sec:fiducial}). Similarly, dust formation is limited at high and low temperatures, as discussed earlier in this section.

Even in the region of parameter space where the outflows are dusty, the production of dust remains inefficient with the dust-to-gas ratio below 0.1 and typically around 0.01. This means that while the total mass-loss rate can easily exceed the $\sim1\,M_\oplus\,{\rm Gyr}^{-1}$ estimated from observations \citep[e.g.][]{Rappaport2012,Sanchis-Ojeda2015, Schlawin2021}, the amount of dust produced in the models is typically less than this. {This is problematic since the aforementioned observational estimates refer to the mass-loss rate of dust.} However, we suspect that the simplifications made here, i.e. homogeneous condensation and grey dust opacities, lead to us underestimate dust production (which is at most $\sim 0.1\,M_\oplus\,{\rm Gyr}^{-1}$). This is because, as described in \autoref{sec:composition}, iron-free silicates would obtain low temperatures and therefore may condense further, producing higher dust-to-gas ratios than is possible in our model.

\subsection{Unsteady flows}
\label{sec:unsteady}
\begin{figure}
	\includegraphics[width=\columnwidth]{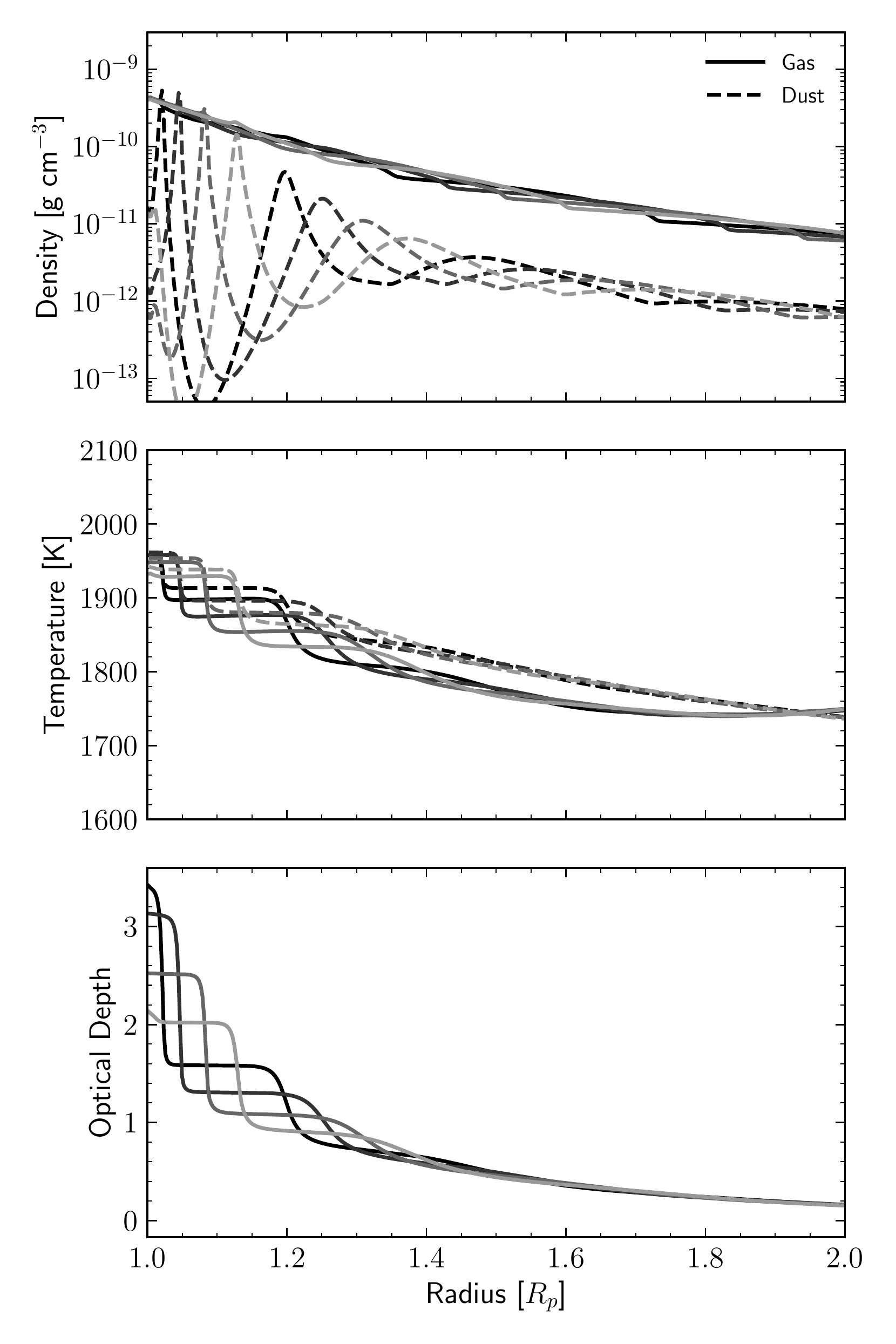}
    \caption{Density and temperature evolution in a model showing unsteady behaviour. The different shades show the structure at different times, each separated by 200~s and sorted in time from dark to light. {The bottom panel shows the optical depth at 1~\micron{} (which is dominated by the dust opacity), corresponding approximately to where most heating and cooling occurs.}}
    \label{fig:unsteady}
\end{figure}

\begin{figure}
	\includegraphics[width=\columnwidth]{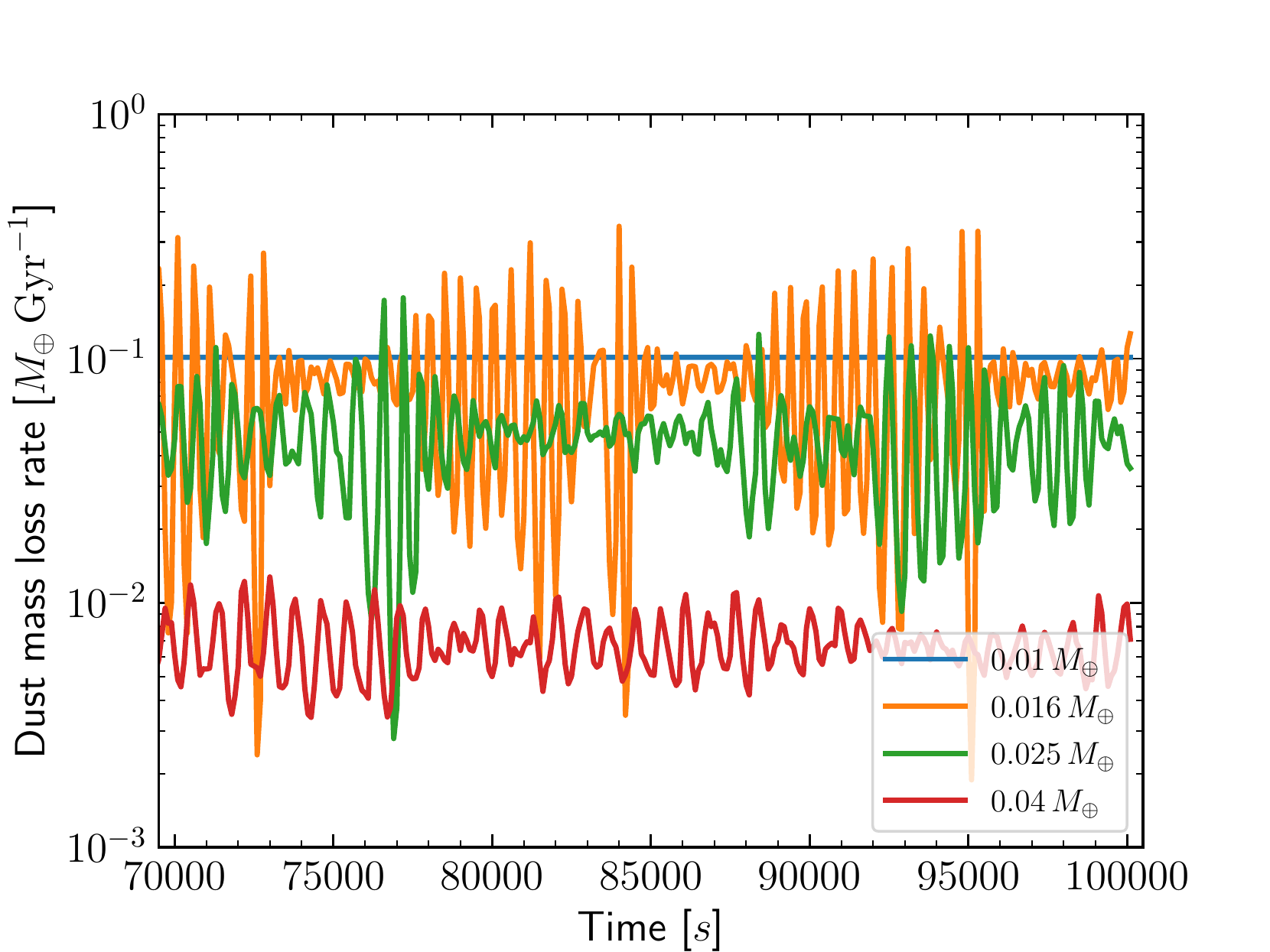}
    \caption{Time variation of the dust mass-loss rate (as measured at the Hill radius) for models with sticking probabilities of $\alpha=1$. These simulations show unsteady behaviour, as discussed in \autoref{sec:unsteady}. } 
    \label{fig:mdot_evo}
\end{figure}

The models presented so far have all reached a steady-state within $10^5$~s (roughly the planet's orbital period, or 10 sound crossing times); however, we found that a subset of models showed variations in their flow properties on time-scales of $\sim 10^3\,{\rm s}$, or roughly the thermal time of the planet's surface (see \autoref{sec:boundaries}). {We confirmed that the time-scale of these variations is related to the planet's surface time-scale by varying the thickness of the layer used in the surface model, $\Delta r$}. These variations did not disappear even after $10^6$~s, much longer than the time taken for the other simulations to reach steady state. Instead they cycled between periods of dust production and no dust production at the planet's surface. An example of such an unsteady flow is shown in \autoref{fig:unsteady}, where the planet mass was $0.02\,M_\oplus$, semi-major axis was 0.015~au, and the sticking probability was $\alpha=1$. While the non-steady flows typically show only small fluctuation in the gas density and mass-loss rate, the amount of dust formed can {vary more substantially, by factors from 1.5 or 2 up to an order of magnitude (\autoref{fig:mdot_evo}). The amplitude of the $10^3\,{\rm s}$ variation itself also varies on a time-scale comparable to the flow time-scale ($\sim 10^4\,{\rm s}$), {although averages of the mass-loss rate on time-scales of $\sim 10^4$ or longer are fairly steady.}}

{It should be noted that the variations in optical depth can be smaller than the variations in mass-loss rate (\autoref{fig:unsteady}) because the flow time-scale is somewhat longer than the time-scale for the variations, causing them to be partially averaged out. The short time-scale for variations implies that the depth of consecutive transits would be unrelated, as seen for example in Kepler-1520 \citep{Rappaport2012, vanWerkhoven2014}. The transit-to-transit variations in depth seen in \citet{vanWerkhoven2014} were of up to a factor $\sim 10$. We would need to couple our simulations to a model for the dynamics of the dusty tail (similar to those of \citealt{vanLieshout2014}, for example) to properly assess whether our models reproduce the magnitude of variations observed, but it is likely that the magnitude of variations seen in our model do not yet fully explain those observed. As discussed later, multi-dimensional simulations are likely to be needed to achieve the necessary degree of realism.}

We found that non-steady flows typically arose in conditions with: 1) fast dust growth rates and 2) moderate optical depths. This favours lower mass planets and larger sticking probabilities. We suggest that these conditions lead to variability due to the following argument: fast dust growth is critical because it leads to dust forming clumps rapidly near the planet's surface. These clumps are then carried outwards by the wind. As the clumps travel outwards the optical depth to the planet's surface decreases, causing the planet's temperature to rise. This increases the difference between the planet's and dust's temperature, eventually causing the conditions near the planet to favour dust formation. Dust forms rapidly once these conditions are met. This increases the optical depth to the planet, causing its temperature to cool, and shutting off dust formation until the newly formed dust is carried out by the wind. 

We suspect that the details of the variability are likely to be affected by the idealised geometry. For example, in 1D the rapid formation of dust clumps leads to the formation of narrow shells of dust, which may be unrealistic. Nevertheless, the main driver for variability is the coupling of the planet's surface temperature to the dust's optical depth, which likely remains true in more realistic scenarios. We note that this idea is essentially the same as proposed by \citet{PB2013}, although they were not able to see such variability in their models since: 1) they did not couple the dust formation prescription to the flow properties and 2) their models assumed steady-state.

\section{Discussion}
\label{sec:discussion}

We have presented the first simulations of outflows from the day sides of ultra-short period planets that couple dynamics, radiative transfer, and dust formation. The model is 1D and applied to outflows from the sub-stellar point. Our models show that dust formation in these outflows is robust, arising naturally when the dust entrained in the winds obtains lower temperatures than the planet's surface. The range of parameter space over which we find dusty outflows are is however limited. To a certain extent, this range is affected by our model assumptions: at low planet masses and temperatures, it is the long growth time-scale of micron-sized grains and the ability of the outflow to entrain those grains that limits dust formation. {Since both of these problems are reduced for sub-micron grains, it is possible that there could be systems in which the outflows contain predominately small grains. There is no evidence for systems with large amounts of sub-micron grains among the three systems known so far, however \citep{Brogi2012, Budaj2013, Sanchis-Ojeda2015, Schlawin2021}.}

One requirement for dust formation is that the dust -- and gas -- temperature must drop below the planet's surface temperature. The dust temperature must fall below the planet's temperature because the gas pressure must exceed the vapor pressure of the dust for condensation to overcome evaporation and the gas pressure is at most equal to the vapor pressure at the planet's surface temperature. However, the gas temperature must also fall below the planet's surface temperature for seed particles to nucleate. This requirement has not always been satisfied in previous models, e.g., \citet{Ito2015} found that the gas temperatures exceeded the planet's surface temperature and therefore argued against dust formation. More recent calculations by \citet{Zilinskas2022} have however found a range of results with the gas temperature sometimes exceeding the planet's temperature, but sometimes not. A key difference between the work of \citet{Zilinskas2022} and \citet{Ito2015} is that the more recent calculations include the  MgO in the opacity, which provides a large opacity in the infrared, enhancing the cooling rates compared with previous calculations. Since we also include MgO as an opacity source, our atmospheres are cooler than those of \citet{Ito2015}. 

\citet{Zilinskas2022} showed that the composition of the atmosphere is one of the key parameters that determines whether the atmospheres are hotter than the planet surface or not, another being the effective temperature. Here we have considered a composition dominated by SiO, Mg, O, etc (\autoref{tab:abund}), which \citet{Schaefer2009} show is appropriate for a planet that has lost a significant amount ($\gtrsim 1$~per cent) of its mass. Atmospheres of planets that have lost little or none of their mass instead would have Na as the main component -- these planets should have dust-free winds on their day sides because Na-dominated compositions have atmospheres that are hotter than their surfaces \citep{Zilinskas2022}. This result is completely consistent with our finding that micron-sized dust forms most readily for planets that would be expected to lose all of their mass within the next $\sim0.1$~Gyr (\autoref{fig:mass_loss}); however, it assumes that composition of the entire planet is affected uniformly by mass loss. Flows from the day-side to the night-side, or inefficient mixing during mass loss could change this picture. If this affects the properties of (or the formation of) dust, then in principle a detailed characterization of the mass loss from these planets might be able to probe these processes.

Our simulations are also the first models for winds from ultra-short period planets that show non-steady behaviour (\autoref{sec:unsteady}). We found that non-steady behaviour arose under conditions associated with fast dust growth and moderate optical depths. Rapid dust formation leads to changes in optical depth that affect the planet's surface temperature, in turn affecting dust formation. In our 1D calculations this produces dust shells that are then carried out by the wind. We suspect that in reality this process may lead to the formation of small dust clumps rather than coherent shells, which would likely affect the way that the variability manifests. Clearly, multi-dimensional simulations are needed to address this. 

Further, multi-dimensional simulations are wanted even in the steady regime due to the strong day-night temperature differences that will drive strong flows to the night side. Although day-night flows have been considered using a shallow water model \citep{Castan2011,Nguyen2020,Kang2021}, these have made simple assumptions about the temperature structure. Further, the atmospheric scale height is comparable to planet's radius under conditions that produce strong outflows, making it unclear if the day-night flows can be modelled as shallow. As with our model, these clearly need to be tested with multi-dimensional simulations.

Finally, we have adopted simple assumptions for the atmospheric chemistry and applied a simplified model for dust growth. We already argued that condensation is likely to be heterogeneous (\autoref{sec:composition}), and it is also possible that heterogeneous condensation may be needed to explain the observations, {but this is something we have not included in our simulations.} Heterogeneous compositions could also partially explain the unexpected results obtained from models of the dusty tails. Specifically, that corundum (Al$_2$O$_3$) best explains their properties \citep{vanLieshout2014} but is not expected to be present in the winds or atmospheres of ultra-short period planets (except perhaps during a final phase when the planets have lost most of their mass). It is possible that composite grains could provide a better match than the pure materials considered so far. 

\section{Conclusions}
\label{sec:conclusions}

We have conducted the first radiation-hydrodynamic simulations of mass loss from the day side of ultra-short period planets. {We have included a simple model for the growth and destruction of dust grains based on the rate at which grains grow or evaporate. Although we did not explicitly model the nucleation of dust grains, our models show that the gas is supersaturated (as also found by \citealt{PB2013}) and therefore seed nuclei are expected to form.} 

Our model demonstrates that these winds are expected to be dusty under a reasonable range of conditions. This process is therefore likely to be responsible for the dust tails seen in a few sources discovered by \emph{Kepler} and \emph{K2} \citep{Rappaport2012,Rappaport2014,Sanchis-Ojeda2015}. 

The reason that dust may form in these winds is that large dust grains with a grey opacity can obtain temperatures that are lower than the planet's surface temperature. The low dust temperatures means that the grains evaporate more slowly than the planet's surface, resulting in net condensation of onto their surfaces. The mass-loss process for gas is similar to previous results \citep[e.g.][]{PB2013}, except we find that heating due to the stellar irradiation and radiative cooling dominate over all other heating and cooling processes, such as latent heat release, adiabatic expansion and dust-gas collisional coupling. Since radiative heating dominates, this allows the gas and dust to obtain temperatures that differ from each other in optically thin regions of the flow. This allows the dust to remain relatively cool ($\sim 1600$~K) even when the gas is hot ($\gtrsim 2500$~K), {and therefore the dust can survive much longer than it would if had the same temperature as the gas.}

In general, dust formation predominantly occurs near the planet's surface instead of far out in the flow. The reason for this is simply that the higher gas pressure leads to faster dust growth rates near the planet. In many cases the lower pressures further away from the planet means that the conditions change from favouring growth to destruction of the grains. This is consistent with canonical picture that the length of the dusty tails is set by dust sublimation \citep[e.g][]{vanLieshout2014}. Nevertheless, we have found conditions where the escaping dust may predominantly form further away from the planet, particularly for lower surface temperatures ($\lesssim 2000$~K). Such conditions arise when the gas densities are low, such that micron-sized grains that form close to the planet cannot be carried out by the wind. However, the low densities far from the planet means that growth times are long, and as a result any grains that form are small. {Such grains are unlikely to be responsible for the tails observed so far however, which favour micron-sized grains \citep[e.g.][]{Schlawin2021}.}

Finally, the conditions in which dust forms corresponds to small planets with high mass loss rates. Therefore, these planets are in the last $\sim 0.1$~Gyr of their existence and the dusty tails we observe signify the end of their evolution, {as also hypothesised by \citet{Rappaport2012} and \citet{PB2013}}.

\section*{Acknowledgements}

We thank Timmy Delage and Beatriz Campos Estrada for interesting conversations, and an anonymous reviewer for constructive criticism. RAB and JEO are supported by the Royal Society through University Research Fellowships. This project has received funding from the European Research Council (ERC) under the European Union’s Horizon 2020 research and innovation programme (Grant agreement No. 853022, PEVAP) and a 2020 Royal Society Enhancement Award. For the purpose of open access, the authors have applied a Creative Commons Attribution (CC-BY) licence to any Author Accepted Manuscript version arising.

\section*{Data Availability}

The simulation code and scripts used to generate these models, along with a grid of mass-loss rates, are available through Zenodo, \href{https://www.doi.org/10.5281/zenodo.7252282}{DOI:10.5281/zenodo.7252282}. The simulation results themselves will be shared on reasonable request to the corresponding author.



\bibliographystyle{mnras}
\bibliography{evap} 




\appendix

\section{Gas Opacity Model}
\label{app:Gas_Opac}

\begin{figure*}
	\includegraphics[width=\textwidth]{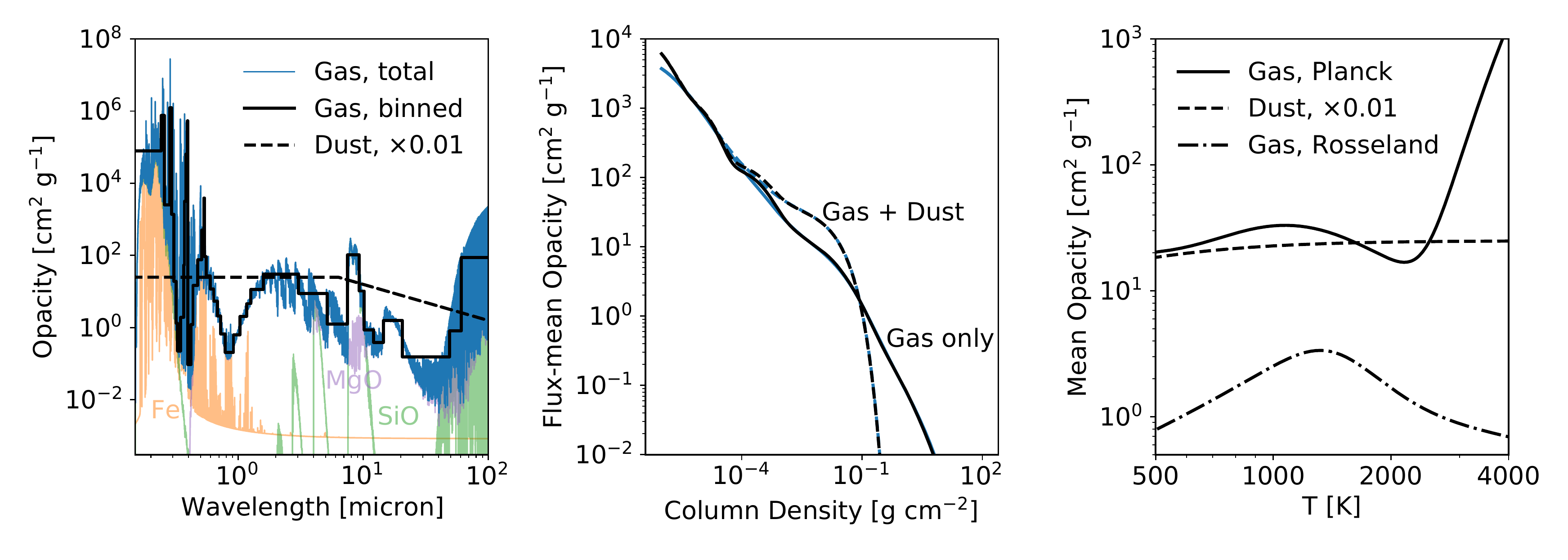}
    \caption{Opacity used in the models. {\bf Left}: the total opacity due to the gas (solid) and dust (dashed). The blue line shows the input spectrum at 2048 points per decade in wavelength, while the black line shows the final binned spectrum (41 wavelength points). Additionally, we show the contribution from the most important species. For the dust opacity a grain size of 1~\micron{} and a dust-to-gas mass ratio of 0.01 were used. {\bf Middle}: The flux-mean opacity for the gas and total (gas + dust) opacity for a 6000~K black-body attenuated by a given column. This shows that the binned spectrum provides a reasonable estimate of the stellar heating rate. {\bf Right}: the Planck and Rosseland mean opacities used. Note that these are assumed to be equal for the dust.}
    \label{fig:opac}
\end{figure*}

We wish to avoid solving an expensive line-by-line radiative transfer problem and therefore use a hybrid model where we use frequency dependent ray-tracing for the heating by stellar irradiation but treat the heating and cooling in the flux-limited diffusion approximation (FLD), as outlined in \autoref{sec:heat_cool}. We therefore need the Planck- and Rosseland-mean opacities for the FLD calculation, along with the frequency-dependent opacity computed in appropriate bins for the stellar heating. 

To compute the frequency-dependent opacity model we start by computing the opacity between 0.01 and 100~\micron{} using 8192 bins logarithmically spaced in wavelength. We assume LTE level populations and include both Doppler and natural broadening via the Voigt profile.  Pressure-broadening is neglected due to the low pressures considered. We include SiO, MgO, Fe, O, and Mg as opacity sources using the line-lists from ExoMol \citep{Li2019,Yurchenko2021} and \citet{Kurucz1992}. Other species present, such as O$_2$ and SiO$_2$ do not contribute significantly to the total opacity \citep[e.g.][]{Ito2015}. The abundances used are given in \autoref{tab:abund}.

For the stellar heating we assume a gas temperature of 2000~K to compute the initial input opacity. We first divide the spectrum into coarse bins by stepping upwards in wavelength and making sure that: i) the Planck function varies by less than a factor 2 in the bin and ii) the maximum wavelength in each coarse bin is less than 1.5 times the minimum one. We then sort the opacity in each of these coarse bins. This sorting procedure introduces a modest 0.1 per cent error into the Planck-mean opacity while simplifying the subsequent steps in the binning procedure. We then proceed by merging the two neighbouring bins that introduces the smallest maximum relative error into the flux-mean opacity for columns between $10^{-5}$ and $10^{3}\,{\rm g\,cm}^2$, assuming a 6000~K black-body for the spectrum\footnote{I.e. $\int {\rm d}\nu\,B_\nu (T) \kappa_\nu \exp{(-\kappa_\nu\Sigma)}/\sigma T^4$, where $\Sigma$ is the column.}. These two bins are then replaced by a single bin with an opacity equal to the Planck-mean of the two over the appropriate wavelength range. This procedure is then repeated iteratively until the desired number of bins (32) is achieved. Since this procedure resulted in an opacity model with just one bin covering wavelengths above $3\,\micron$, we then filled in the long-wavelength region by replacing the last bin with a spectrum computed with twice the number of bins. After making such a replacement 4 times, we were arrive at the opacity model shown in \autoref{fig:opac} (left panel), which has 41 bins in total. The middle panel shows that this model successfully reproduces the flux-mean opacity over a wide range of column densities, and therefore also reproduces the stellar heating rate, as required. We note that simply tabulating the stellar heating rate would be possible, but would require a 3D table due to the significant, and variable, contribution to the opacity from the dust grains.

Finally, we need the Planck-mean and Rosseland-mean opacities of the gas. To compute the Planck-mean opacity, we simply computed the LTE level populations and added the contribution of individual lines neglecting any line broadening. Tests of a few specific cases showed that this works sufficiently well. A simple model of the Rosseland-mean opacity from the gas is sufficient because the Rosseland-mean opacity is dominated by the dust grains. For this reason we computed the Rosseland mean of the 8192 bin spectrum that was used as input for the stellar heating calculation at different temperatures. This, and the other mean opacities used, are shown in the right panel of \autoref{fig:opac}.

\begin{table}
	\centering
	\caption{Assumed abundances. Based on \citet{Schaefer2009}.}
	\label{tab:abund}
	\begin{tabular}{lc} 
		\hline
		Species & Mole Fraction\\
		\hline
		SiO & 0.281 \\
		Mg & 0.250 \\
		O & 0.223 \\
		O$_2$ & 0.158 \\
		Fe  & 0.079 \\
		SiO$_2$ & 0.005 \\
		MgO & 0.003 \\
		\hline
	\end{tabular}
\end{table}

\section{Up and Downwards fluxes in the flux-limited diffusion approximation}
\label{app:FLD_updown}
The amount of radiation passing upwards and downwards from a given point are not immediately available in the flux-limited diffusion approximation, which only determines the net flux. To split this flux into its upgoing and downgoing contributions we need a model for the angular dependence of the intensity, $I$. For this we use an extension of the Eddington approximation, $I(\mu) = A + B\mu + C\delta(\mu-1)$, where $\mu$ is the cosine of the angle and $\mu = 1$ refers to the direction away from the planet. The case $C=0$ is the normal Eddington approximation corresponding to the diffusion limit \citep[e.g.][]{MihalasMihalas}, while $I(\mu) = C\delta(\mu-1)$ allows for the free-streaming limit. With this approximation we have:
\begin{align}
 J &= \frac{1}{2}\int_{-1}^1 I(\mu) \diff{\mu} = A + \frac{1}{2}C \\ 
 H &= \frac{1}{2}\int_{-1}^1 I(\mu) \mu \diff{\mu} = \frac{1}{3} B + \frac{1}{2}C \\ 
 K &= \frac{1}{2}\int_{-1}^1 I(\mu) \mu^2 \diff{\mu} = \frac{1}{3}A + \frac{1}{2}C \\ 
 H_\uparrow &= \frac{1}{2}\int_{0}^1 I(\mu) \mu \diff{\mu} = \frac{1}{4}A + \frac{1}{6} B + \frac{1}{2}C \\ 
 H_\downarrow &= \frac{1}{2}\int_{0}^{-1} I(\mu) \mu \diff{\mu} = \frac{1}{4}A - \frac{1}{6} B
\end{align}
We then re-write $ H_\uparrow$ and $ H_\downarrow$ in terms of $J$, $H$, and $K$:
\begin{align}
 H_\uparrow &= \frac{1}{8}\left(J + 3K\right) + \frac{1}{2}H \\
 H_\downarrow &= \frac{1}{8}\left(J + 3K\right) - \frac{1}{2}H
\end{align}
The standard flux-limited diffusion equations provide $4\upi H = F$ in terms of $J$ (\autoref{eqn:FLD_flux}), but we still need an expression for $K$. For this we follow the recommendation of \citet{Levermore1984} and use the closure
\begin{align}
 \frac{K}{J}=  \lambda(J) + \left[\lambda(J) \mathcal{R}(J)\right]^2
\end{align}
where $\lambda(J)$ is the flux-limiter and $\mathcal{R}(J) = |\kappa\nabla J/J|$. In the free-streaming limit $\lambda(J) \rightarrow 1/\mathcal{R}(J)$ as $\mathcal{R}(J) \rightarrow \infty$, from which it is straight-forward to demonstrate that  $H_\downarrow \rightarrow 0$ and  $H_\uparrow \rightarrow J$ as required. Similarly, the usual results for the diffusion limit ($\mathcal{R}(J) \rightarrow 0$, $\lambda(J) \rightarrow 1/3$) are recovered. Finally, the relations $F_\uparrow = 4 \upi H_\uparrow$  and  $F_\downarrow = 4 \upi H_\downarrow$ provide the required up and downwards fluxes for the boundary conditions.

\section{Source term integration}
\label{app:numerics}

Here we briefly describe the full method used in \textsc{aiolos} \citep{Schulik2022} to couple the source terms associated with dust condensation and sublimation with the hydrodynamics and radiative transport. The code operates by cycling through each physics sub-step using operator-splitting. First the hydrodynamics step is done and the drag forces are applied. During this step the frictional heating due to drag, $\dot{\mathcal{E}}_{1,s}$, is applied, but the collisional heating term, $\dot{\mathcal{E}}_{2,s}$, is deferred. 

Next we solve the source terms due to condensation and sublimation. During this step we compute the exchange of mass and momentum along with the net heating rate associated. The update is done implicitly, including the radiative heating and cooling, $\Gamma_s$, and collisional heat exchange terms, $\dot{\mathcal{E}}_{2,s}$. We hold the thermal radiation, $J$, fixed during this step, but update the stellar irradiation terms since the density and optical depth change. For completeness, the equations solved during this step are 
\begin{align}
    \pderiv{\rho_s}{t} &= Q_s,  \\
    \pderiv{\rho_s u_s}{t} &= Q'_s, \\
    \pderiv{E_s}{t} &= Q''_s + \dot{\mathcal{E}}_{2,s}  + \Gamma_s. \label{app:eqn:T_eqn}
\end{align}
The inclusion of the radiative heating and collisional terms in this step ensures that we can correctly capture cases where condensation/sublimation is limited by heating or cooling of the grains, without requiring time-steps smaller than the condensation/sublimation time-scale. We use the backward-Euler method to solve these equations for the new density, momentum, and temperature of the gas and dust. Once this is done we immediately update the density and momentum, but the temperature update is deferred. 

Finally, we re-solve \autoref{app:eqn:T_eqn} along with the radiative transfer equation, \autoref{eqn:FLD}, to get the final temperature. In this step we are careful to use exactly the same heating rates ($Q''_s$, stellar heating, etc.) computed during the condensation/sublimation step to ensure energy conservation. Note that the heating is not double-counted because the result of the previous evaluation of \autoref{app:eqn:T_eqn} is discarded -- i.e. it is only the heating and cooling rates that are kept, not the updated energies.

Now only the update of the planet's surface temperature remains as described in \autoref{sec:boundaries}. This is again done implicitly, but with the stellar heating and thermal flux at the planet's boundary held constant using the values determined in the previous step. This completes the time-step.


\bsp	
\label{lastpage}
\end{document}